\documentclass{elsarticle}

\usepackage{subfig}
\usepackage{amssymb,amsmath,mathrsfs}%
\usepackage{gensymb}
\usepackage{amsthm,bm}%
\usepackage[width=17cm,height=23.5cm,centering]{geometry}
\usepackage{color}
\usepackage{tcolorbox} %
\usepackage{stmaryrd} 
\usepackage{cases,empheq,algorithm,enumitem}
\usepackage{todonotes} 
\usepackage{soul}
\usepackage{cancel}

\usepackage[scr=boondox]{mathalfa}

\newcommand{\ds}{\displaystyle}

\newcommand{\jump}[1]{\left\llbracket #1 \right\rrbracket}
\newcommand{\mean}[1]{\left\langle #1 \right\rangle}
\newcommand{\dmean}[1]{\left\llangle #1 \right\rrangle}

\makeatletter
\DeclareFontFamily{OMX}{MnSymbolE}{}
\DeclareSymbolFont{MnLargeSymbols}{OMX}{MnSymbolE}{m}{n}
\SetSymbolFont{MnLargeSymbols}{bold}{OMX}{MnSymbolE}{b}{n}
\DeclareFontShape{OMX}{MnSymbolE}{m}{n}{
    <-6>  MnSymbolE5
   <6-7>  MnSymbolE6
   <7-8>  MnSymbolE7
   <8-9>  MnSymbolE8
   <9-10> MnSymbolE9
  <10-12> MnSymbolE10
  <12->   MnSymbolE12
}{}
\DeclareFontShape{OMX}{MnSymbolE}{b}{n}{
    <-6>  MnSymbolE-Bold5
   <6-7>  MnSymbolE-Bold6
   <7-8>  MnSymbolE-Bold7
   <8-9>  MnSymbolE-Bold8
   <9-10> MnSymbolE-Bold9
  <10-12> MnSymbolE-Bold10
  <12->   MnSymbolE-Bold12
}{}

\let\llangle\@undefined
\let\rrangle\@undefined
\DeclareMathDelimiter{\llangle}{\mathopen}%
                     {MnLargeSymbols}{'164}{MnLargeSymbols}{'164}
\DeclareMathDelimiter{\rrangle}{\mathclose}%
                     {MnLargeSymbols}{'171}{MnLargeSymbols}{'171}
\makeatother

\definecolor{rouge}{rgb}{1,0,0}
\definecolor{bleu}{rgb}{0,0,1}
\definecolor{vert}{rgb}{0,0.5,0}

\newtheorem{Lemma}{Lemma}
\newtheorem{Proposition}{Proposition}

\newtheorem{Remark}{Remark}

\graphicspath{{Figures/}}

\begin{document}
 \begin{frontmatter}
\title{Waves within a network of slowly time-modulated interfaces: time-dependent effective properties, reciprocity and high-order dispersion}

\author[LMA]{Micha\"el Darche}
\ead{darche@lma.cnrs-mrs.fr}
\author[UM]{Raphael Assier\corref{cor1}}
\ead{raphael.assier@manchester.ac.uk}
\author[UMI-ICL,Phys-ICL]{S\'ebastien Guenneau}
\ead{s.guenneau@imperial.ac.uk}
\author[LMA]{Bruno Lombard}
\ead{lombard@lma.cnrs-mrs.fr}
\author[Math-ICL,UMI-ICL,POEMS]{Marie Touboul}
\ead{marie.touboul@ensta.fr}
\cortext[cor1]{Corresponding author}
\address[LMA]{Aix Marseille Université, CNRS, Centrale Marseille, LMA UMR 7031, Marseille, France}
\address[UM]{Department of Mathematics, University of Manchester, Oxford Road, Manchester M13 9PL, UK}
\address[UMI-ICL]{UMI 2004 Abraham de Moivre-CNRS, Imperial College London, London SW7 2AZ, UK}
\address[Phys-ICL]{Department of Physics, The Blackett Laboratory, Imperial College London, London SW7 2AZ, UK}
\address[Math-ICL]{Department of Mathematics, Imperial College London, Huxley Building, Queen's Gate, London SW7 2AZ, UK}
\address[POEMS]{POEMS, ENSTA Paris, CNRS, INRIA, Institut Polytechnique de Paris, 91120, Palaiseau, France}
%


\begin{keyword}
Time-varying media, high-order homogenisation, $k$-gap, reciprocity, time-domain numerical methods\\ \textit{Subjects: }wave motion, applied mathematics, acoustics, asymptotic methods
\end{keyword}

\begin{abstract}
We consider wave propagation through a 1D periodic network of slowly time-modulated interfaces. Each interface is modelled by time-dependent spring-mass jump conditions, where mass and rigidity interface parameters are modulated in time. Low-frequency homogenisation yields a leading-order model described by an effective time-dependent wave equation, i.e.\ a wave equation with effective mass density and Young's modulus which are homogeneous in space but depend on time. This means that time-dependent bulk effective properties can be created by an array where only interfaces are modulated in time. The occurrence of k-gaps in case of a periodic modulation is also analysed. Second-order homogenisation is then performed and leads to an  effective model which is reciprocal but encapsulates higher-order dispersive effects. These findings and the limitations of the models are illustrated through time-domain simulations. 
\end{abstract}
 \end{frontmatter}

\section{Introduction}
The spatial modulation of material properties has, since the early 2000s, enabled progress in wave manipulation, with the emergence of \textit{metamaterials}. We refer the reader to \cite{craster2013,craster2023mechanical} for an overview of the subject in acoustics and mechanics. More recently, the idea of varying properties both in space and time, which dates back to the pioneering works \cite{Simon1960,Oliner1961,Cassedy1963,Cassedy1967}, has gained renewed attention \cite{caloz2019spacetime1,caloz2019spacetime2,galiffiA2022}. Indeed, interaction of waves with media whose properties are space-time dependent offer additional degrees of freedom for the control of waves compared to their spatial counterparts such as nontrivial topology \cite{Swinteck2015}, parametric amplification \cite{cullen1958travelling,tien1958parametric}, unidirectional amplification \cite{wenCP2022}, coherent perfect absorption \cite{Galiffi2026} and non-reciprocity \cite{Croenne2019,Nassar2020,tessierbrothelandeAPL2023} without relying on non-linearities or on the presence of a flow.  

Despite these advances, significant challenges remain for the experimental realisation of temporal modulation of bulk parameters such as density or bulk modulus. In practice, time modulation is more easily achieved at discrete points, for example by tuning the stiffness of a membrane or a surface impedance \cite{zhuPRB2020,zhuAPL2020}, or by vertical oscillations of a submerged plate \cite{Koukouraki2025}. To analyse such configurations, several approaches have been developed, including multiple scattering \cite{puJoSaV2024}, transmission-line theory \cite{mallejacPRA2023}, and capacitance-matrix formulations \cite{ammari_scattering_2024,ammari_spacetime_2025}. In a recent work, we have studied theoretically and numerically the scattering of waves by one time-modulated interface, and the associated generation of Floquet harmonics \cite{Darche2025}. 

In this article, motivated by the metamaterials paradigm, where the repetition of subwavelength features makes possible macroscopic wave control, we extend the study \cite{Darche2025} to periodic networks of time-modulated interfaces. This configuration is similar to the one investigated experimentally in \cite{tessierbrothelandeAPL2023}, and also to the periodic settings of resonators studied theoretically in \cite{Ammari2021,ammariJoMP2023,Ammari2022}. To this end, we use the two-scale asymptotic expansion method  \cite{Bensoussan2011,Bakhvalov1989,cioranescu1999,Conca1997,Boutin1995} to homogenise in space the considered medium. Our contribution here is then to provide an effective medium approximation up to second order, which sheds light on the behaviour of waves in an array of time-modulated interfaces. Our approach differs from homogenisation of time-dependent {\it bulk} media, investigated in various works. One can cite e.g.\ laminates with wave-like modulations of two parameters leading to Willis-coupling in acoustics \cite{Nassar2017,Touboul2024} and magneto-electric coupling in electromagnetics \cite{Huidobro2019,Huidobro2021}.  A leading-order asymptotic analysis of the low-frequency regime for the time-modulated wave equation is also addressed in \cite{lurie2007introduction,to2009homogenization,sanguinet2011homogenized}. 

We start by presenting the setting of a periodic network of time-modulated interfaces in Section~\ref{Sec:Setting}. Leading-order homogenisation is then performed in Section~\ref{SecLeading} and leads to an effective wave equation with time-dependent effective coefficients. The properties of this model are analysed, including the emergence of wavenumber gaps when the modulation is periodic. High-order homogenisation, up to the second-order, is then performed in Section~\ref{Sec:HomogOrdre2} and we prove that this effective model is always reciprocal. Finally, time-domain simulations are presented in Section~\ref{Sec:Num}, providing validation of the models and illustrating the behaviour of the effective models, as well as the limits of their validity.


\section{Setting: periodic network of time-modulated interfaces}\label{Sec:Setting}
\vspace{-15pt}
\begin{figure}[htbp]
\begin{center}
\begin{tabular}{c}
\includegraphics[scale=0.5]{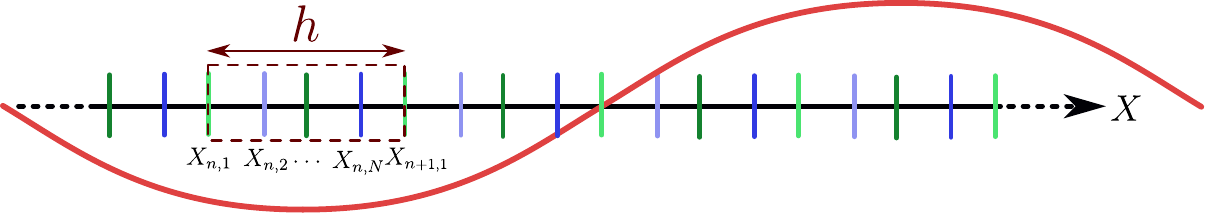} 
\end{tabular}
\end{center}
\caption{\label{FigScales}Periodic network of $N=4$ imperfect interfaces located at $X_{n,\ell}$ and repeated periodically with a period $h$. A typical wavelength is represented in red.}
\end{figure}
\vspace{-10pt}
We consider the propagation of transient waves in a 1D $h$-periodic linear elastic medium of mass density $\rho_h(X)=\rho(X/h)$ and Young's modulus $E_h(X)=E(X/h)$, where $\rho$ and $E$ are 1-periodic strictly positive functions {and $X$ is the dimensional space variable}. The medium contains a periodic network of imperfect interfaces, see Figure \ref{FigScales}. More precisely, we introduce the periodic cells $(nh,\,(n+1)h)$, with $n\in \mathbb{Z}$. Within each cell, $N$ interfaces are repeated periodically and located at $X_{n,\ell}$, with $\ell=1,\cdots,N$ denoting the index of the interface and $n\in\mathbb{Z}$ the periodic cell considered. {Without loss of generality we may assume that $X_{0,1}=0$.} The interfaces $X_{n,1}$  coincide with the boundaries of the periodic cells. If $N>1$, the interfaces are not necessarily uniformly distributed within the cell. 

Given a source term $F$, the conservation of momentum and the Hooke's law result in
\begin{subnumcases}{\label{EDPmicro}}
\ds \rho_h(X)\,\partial^2_{TT}U_h(X,T) =\partial_X S_h(X,T)+F(X,T),\label{EDPmicro-a}\\ [-6pt]
\ds S_h(X,T) = E_h(X)\,\partial_XU_h(X,T),\label{EDPmicro-b}
\end{subnumcases}
with $U_h$ the displacement field, $S_h$ the stress field, and $T$ the dimensional time variable. The system \eqref{EDPmicro} is completed by initial conditions. Both $U_h$ and $S_h$ are continuous on the open intervals $(X_{n,1},X_{n+1,1})$, except at inner interfaces $X_{n,\ell}$. 

For any function $g(X)$, we define the jump and mean operators across the interface $X_{n,\ell}$:
\begin{equation}
\label{def_jump_mean_dim}
\jump{g}_{X_{n,\ell}} = g^+(X_{n,\ell})-g^-(X_{n,\ell}), \qquad \dmean{g}_{X_{n,\ell}}=\frac{1}{2}\left(g^+(X_{n,\ell})+g^-(X_{n,\ell})\right),
\end{equation}
where 
\begin{equation}
\label{def_gpm}
g^{\pm}(X_{n,\ell}) = \lim_{{\chi}\to 0^+}g(X_{n,\ell}\pm{\chi}).
\end{equation}
For the $\ell$-th interface within a periodic cell, we introduce its rigidity $\mathscr{K}_\ell>0$, compliance ${\mathscr{C}_\ell=1/\mathscr{K}_\ell}$, mass $\mathscr{M}_\ell\geq 0$, and dissipation $\mathscr{Q}_{C,\ell}\geq 0$ and $\mathscr{Q}_{M,\ell}\geq 0$. Across these interfaces, the fields satisfy the following time-dependent jump conditions: 
\begin{equation}
\begin{array}{l}
\ds \jump{V_h(\cdot, T)}_{X_{n,\ell}}=\partial_T\left(\mathscr{C}_\ell(T)\,\dmean{S_h(\cdot,T)}_{X_{n,\ell}}\right)+\mathscr{Q}_{C,{\ell}}\dmean{S_h(\cdot,T)}_{X_{n,\ell}},\\ [8pt]
\ds \jump{S_h(\cdot,T)}_{X_{n,\ell}}=\partial_T\left(\mathscr{M}_\ell(T)\,\dmean{V_h(\cdot,T)}_{X_{n,\ell}}\right)+\mathscr{Q}_{M,{\ell}}\dmean{V_h(\cdot,T)}_{X_{n,\ell}},
\end{array}
\label{JCmicro}
\end{equation}
where we have introduced the velocity field $V_h(X,T)=\partial_T U_h(X,T)$.

\section{Leading-order homogenisation}\label{SecLeading}

For the sake of simplicity, we will consider here the case without dissipation, i.e.\ $\mathscr{Q}_{C,\ell}=\mathscr{Q}_{M,\ell}= 0$, {which will be investigated up to second order.} The case with dissipation is considered at leading-order in Appendix \ref{Sec:Homog_Dissip}, {involving another formalism}.
\vspace{-5pt}
\subsection{Two-scale analysis}\label{SecHomog2scale}

\paragraph{\bf Non-dimensionalisation.}We consider a reference wavelength $\lambda^\star$ and we define
\begin{equation}
k^\star=\frac{2\pi}{\lambda^\star}\quad \text{ and } \quad
\eta=k^\star\,h.
\label{Eta}
\end{equation}
Arbitrary reference material properties are introduced: $E^\star$, $\rho^\star$ that define the wave speed $c^\star=\sqrt{\frac{E^\star}{\rho^\star}}$. We define the non-dimensionalised space and time variables $x=k^\star X$ and $t=k^\star c^\star T$, respectively, as well as the non-dimensionalised fields and forcing
\begin{equation}
u_\eta(x,t)=k^\star\,U_h(X,T),\qquad \sigma_\eta(x,t)=\frac{1}{E^\star}\,S_h(X,T),\qquad f(x,t)=\frac{1}{k^\star\,E^\star}\,F(X,T).
\label{Adim}
\end{equation}
Upon defining 
\begin{equation}
\alpha=\frac{\rho}{\rho^\star},\quad \beta=\frac{E}{E^\star},
\label{AlphaBeta}
\end{equation}
the non-dimensionalised momentum equation and  constitutive law become
\begin{equation}
\left\{
\begin{array}{l}
\ds \partial_t\left(\alpha\left(\frac{x}{\eta}\right)\partial_tu_\eta(x,t)\right)=\partial_x\sigma_\eta(x,t)+f(x,t),\\ [5pt]
\ds \sigma_\eta(x,t) = \beta\left(\frac{x}{\eta}\right)\partial_xu_\eta(x,t).
\end{array}
\right.
\label{EDPadim}
\end{equation}
Then the time-dependent jump conditions are recast as
\begin{equation}
\ds \jump{u_\eta(\cdot, t)}_{x_{n,\ell}}=\eta\,\mathscr{c}_\ell(t)\,\dmean{\sigma_\eta(\cdot,t)}_{x_{n,\ell}},\qquad \ds \jump{\sigma_\eta(\cdot,t)}_{x_{n,\ell}}=\eta\,\partial_t\left(\mathscr{m}_\ell(t)\,\dmean{\partial_tu_\eta(\cdot,t)}_{x_{n,\ell}}\right),
\label{JCadim}
\end{equation}
with the non-dimensionalised interface parameters
\begin{equation}
\mathscr{k}_\ell(t)=\frac{\mathscr{K}_\ell(T)\,h}{E^\star},\quad \mathscr{m}_\ell(t)=\frac{\mathscr{M}_\ell(T)}{\rho^\star\,h},\quad \mathscr{c}_\ell(t)=1/\mathscr{k}_\ell(t).
\label{KMadim}
\end{equation}
No scaling is considered in the constitutive parameters, so that $\alpha$, $\beta$, $\mathscr{c}_\ell$, $\mathscr{m}_\ell$, $\mathscr{c}'_\ell$ and $\mathscr{m}'_\ell$ are all assumed to be ${\cal O}(1)$ coefficients in the forthcoming asymptotic expansions, with prime notation denoting time derivative. It means that $\mathscr{M}_\ell$ and $\mathscr{K}_\ell$ must be of the same order of magnitude as $\rho^\star h$ and $E^\star/h$, respectively and{, importantly for the present work, should} not vary too quickly.

\vspace{-7pt}
\paragraph{\bf Asymptotic expansion.}We assume that $0<\eta\ll 1$ meaning that we are in the long-wavelength, or low-frequency, regime\footnote{As mentioned in Introduction, Floquet harmonics are generated due to time modulation (see e.g.\ \cite{Darche2025}). As a consequence, there are different possible choices of the reference wavelength $\lambda^\star$. In Section \ref{Sec:Num} dedicated to numerical simulations, we will be more specific regarding the associated non-dimensionalised parameters.}. In this context, the material parameters $\alpha$ and $\beta$ vary on a fine scale associated with the rescaled coordinate $y=x/\eta$. Since $\eta\ll 1$, scale separation occurs, which implies that the small-scale features of the wave fields are described by $y$, whereas the slow continuous variations are described by the variable $x$.

As is customary in two-scale analysis, we now assume that $u_\eta$ depends on both $x$ and $y$, so that $\partial_x$ becomes $\partial_x+\frac{1}{\eta}\partial_y$. The non-dimensionalised system \eqref{EDPadim} is then recast as 
\begin{equation}
\ds \partial_{t}(\alpha(y)\,\partial_{t}u_\eta)=\left(\partial_x+\frac{1}{\eta}\partial_{y}\right)\left[\beta(y)\,\left(\partial_x+\frac{1}{\eta}\partial_{y}\right)u_\eta\right]+f(x,t).
\label{EDPscaled}
\end{equation}
The field $u_\eta$ is expanded using the usual ansatz:
\begin{equation}
u_\eta(x,t) {\equiv u_\eta(x,y,t)}=\sum_{j\geq 0} \eta^j u_j(x,y,t).
\label{ansatz}
\end{equation}
For all $j\geq1$, the field $u_j$ is assumed to be continuous with respect to the first variable and $1$-periodic with respect to the second variable, i.e.~$u_j(x,y,t)=u_j(x,y+1,t)$ for all $y$. The positions of the imperfect interfaces are 
\begin{equation}
y_{n,\ell}=\frac{x_{n,\ell}}{\eta},\qquad n\in\mathbb{Z} \quad \text{and} \quad \ell=1,\cdots, N.
\label{Supercell-y}
\end{equation}
By periodicity, it is enough to consider only the $0$-th periodic cell ${y\in}(0,1)$ and to simply denote the position of the interfaces in this unit cell by $y_\ell$, with $y_1=0$ and $y_{N+1}=1$. The jump and mean notations \eqref{JCadim} are extended at the micro-scale. Given a 1-periodic function $p{(y)}$ with discontinuities at $y_\ell$, one writes for $\ell=2,\cdots, N$,
\begin{equation}
\begin{array}{l}
\ds \jump{p}_\ell=p(y_{\ell}^+)-p(y_{\ell}^-) \,\text{ and } \,\ds \dmean{p}_\ell=\frac{1}{2}\left(p(y_{\ell}^+)+p(y_{\ell}^-)\right).
\end{array}
\label{jump_mean_uj_2}
\end{equation}
For $\ell=1$, the 1-periodicity with respect to $y$ yields
\begin{equation}
\begin{array}{l}
\ds \jump{p}_1 =p(0^+)-p(1^-) \,\text{ and }\,\ds \dmean{p}_1 =\frac{1}{2}\left(p(0^+)+p(1^-)\right).
\end{array}
\label{jump_mean_uj_1}
\end{equation}

{Using these notations, the jump conditions \eqref{JCadim} become} 
\begin{subnumcases}{\label{CascadeJC}}
\ds \jump{u_\eta}_{\ell}=\eta\,\mathscr{c}_\ell(t)\,\dmean{\beta(y)\,\left(\partial_x+\frac{1}{\eta}\partial_{y}\right)u_\eta}_{\ell},\label{CascadeJCu}\\ 
\ds \jump{\beta(y)\,\left(\partial_x+\frac{1}{\eta}\partial_{y}\right)u_\eta}_{\ell}=\eta\,\partial_t\left(\mathscr{m}_\ell(t)\,\dmean{\partial_tu_\eta}_{\ell}\right).\label{CascadeJCS}
\end{subnumcases}
At this stage, one more notation should be introduced for taking the average of a function $p$ over $(0,1)$. We note
\begin{equation}
\mean{p}=\int_0^1p(y)\,dy.
\label{Average}
\end{equation}
This allows a compact formulation of the following important lemma.
\begin{Lemma}
For any two 1-periodic functions $(p,q)$ with discontinuities at $y_\ell$, one has 
\begin{equation}
\mean{\frac{dp}{dy}\,q}=-\mean{p\,\frac{dq}{dy}}-\sum_{\ell=1}^{N}\jump{p\,q}_\ell=-\mean{p\,\frac{dq}{dy}}-\sum_{\ell=1}^{N}\left( \jump{p}_\ell\dmean{q}_\ell+\dmean{p}_\ell\jump{q}_\ell\right).
\label{JClemma}
\end{equation}
\label{LemmaAve}
\end{Lemma}
\vspace{-20pt}
{\noindent The first equality of the Lemma comes from a standard integration by parts, while the second comes from the useful property $\jump{p\,q}_\ell=\jump{p}_\ell\dmean{q}_\ell+\dmean{p}_\ell\jump{q}_\ell$, which can be checked directly.}
\begin{Remark}
Taking $q=1$ in the above lemma leads to 
\begin{equation}
\label{meandy_jump}
\mean{\frac{dp}{dy}}=-\sum_{\ell=1}^{N}\jump{p}_\ell.
\end{equation}
\end{Remark}


\subsection{Leading-order effective model}\label{SecNHomogOrdre0}
\paragraph{\bf Model derivation.}{As is classical in multiple scale homogenisation, one can input the ansatz \eqref{ansatz} into \eqref{EDPscaled} and \eqref{CascadeJC} and collect the relevant asymptotic orders to obtain a cascade of simpler problems.} The ${\cal O}(\eta^{-2})$ terms in \eqref{EDPscaled}, ${\cal O}(\eta^{0})$ terms in \eqref{CascadeJCu}  and  ${\cal O}(\eta^{-1})$ terms in \eqref{CascadeJCS} yield
\begin{subnumcases}{\label{JC_o0}}
\ds \partial_y \left(\beta \partial_y u_0\right)=0 \label{EDP_-2}, \\[-6pt]
\ds \jump{ u_0}_\ell=\mathscr{c}_\ell(t)\,\dmean{\beta\partial_{y}u_0}_{\ell} \label{JCc_0}, \\[-6pt]
\jump{\beta\,\partial_{y}u_0}_{\ell}=0\label{JCm_-1},
\end{subnumcases}
{for $\ell=1,\cdots,N$.} Integrating \eqref{EDP_-2} together with \eqref{JCm_-1} leads to :
\begin{equation}
        \partial_y u_0=\frac{1}{\beta}\mathscr{C}_{\partial u_0}(x,t),
        \label{eq:Raph318}
\end{equation}
{for some unknown function $\mathscr{C}_{\partial u_0}$}. Averaging \eqref{eq:Raph318} and using \eqref{JCc_0} and \eqref{meandy_jump} gives: 
\begin{equation}
    -\sum_{\ell=1}^{N} \mathscr{c}_\ell(t)\mathscr{C}_{\partial u_0}(x,t)=\mean{\frac{1}{\beta}}\mathscr{C}_{\partial u_0}(x,t).
\end{equation}
Since $\mean{\frac{1}{\beta}}$ and $\sum_\ell \mathscr{c}_\ell(t)$ are strictly positive, one gets $\mathscr{C}_{\partial u_0}(x,t)=0$.
Consequently
\begin{equation}
\label{macro_v0_sigma_0}
u_0(x,y,t) = \overline{u}_0(x,t).
\end{equation}
The overline indicates that the fields depend only on the slow space variable. 

Then, the ${\cal O}(\eta^{-1})$ terms in \eqref{EDPscaled}, ${\cal O}(\eta^{1})$ terms in \eqref{CascadeJCu} and ${\cal O}(\eta^{0})$ terms in \eqref{CascadeJCS} yield
\begin{subnumcases}{\label{JC_o1}}
\ds \partial_y \left(\beta \partial_y u_1\right)+\partial_y(\beta)\partial_x\overline{u}_0(x,t)=0 \label{EDP_-1},\\[-6pt]
\ds \jump{ u_1}_\ell=\mathscr{c}_\ell(t)\,\dmean{\beta(y)\left(\partial_{x}u_0+\partial_{y}u_1\right)}_{\ell} \label{JCc_1},\\[-6pt]
\jump{\beta(y)\,\left(\partial_xu_0+\partial_{y}u_1\right)}_{\ell}=0, \label{JCm_0}
\end{subnumcases}
for $\ell=1,\cdots,N$. The governing equation \eqref{EDP_-1} is an ordinary differential equation in $y$ with a source term of the form $a(y)\, b(x,t)$. It ensures that $u_1$ can be written as 
\begin{equation}
\label{form_u1}
u_1(x,y,t)=\overline{u}_1(x,t)+P_1(y,t)\,\partial_x\overline{u}_0(x,t),
\end{equation}
where the \textit{corrector} $P_1$ is solution of the following \textit{cell problem} $(\mathscr{P}_1)$:
\begin{equation}
(\mathscr{P}_1):
\left|
\begin{array}{ll}
\ds \partial_y(\beta(\partial_yP_1+1))=0 , & \text{ on } \bigcup\limits_{\ell=1}^{N}(y_{\ell},y_{\ell+1}) \\[-2pt]
\ds\jump{P_1}_\ell = \mathscr{c}_\ell(t)\, \dmean{\beta(\partial_yP_1+1)} ,\quad   &\text{ for } \ds \ell=1,\cdots,N\\[-2pt]
\ds \jump{\beta(\partial_yP_1+1)}_\ell = 0, & \text{ for } \ds \ell=1,\cdots,N\\[-2pt]
\ds \mean{P_1}=0.\quad P_1\,\mbox{ 1-periodic in }y.&
\end{array}
\right.
\label{pb_cell_P1}
\end{equation}
One notes that $P_1$ depends not only on $y$ (as usual), but varies also with time because of the time-modulation of the interface compliance.

The ${\cal O}(\eta^{0})$ terms in \eqref{EDPscaled}, ${\cal O}(\eta^{2})$ terms in \eqref{CascadeJCu} and ${\cal O}(\eta^{1})$ terms in \eqref{CascadeJCS} yield
\begin{subnumcases}{\label{JC_o2}}
\ds \partial_{t}(\alpha(y)\,\partial_{t} u_0)=\partial_y(\beta\partial_yu_2)+\partial_y(\beta\partial_xu_1)+\beta\partial^2_{xy}u_1+\beta\partial^2_{xx}u_0+f(x,t) \label{EDP_0},\\[-6pt]
\ds \jump{ u_2}_\ell=\mathscr{c}_\ell(t)\,\dmean{\beta(y)\left(\partial_{x}u_1+\partial_{y}u_2\right)}_{\ell} \label{JCc_2},\\[-6pt]
\jump{\beta(y)\,\left(\partial_xu_1+\partial_{y}u_2\right)}_{\ell}=\partial_t\left(\mathscr{m}_\ell(t)\,\dmean{\partial_{t}u_0}_{\ell}\right). \label{JCm_1}
\end{subnumcases}
By averaging the first equation and using \eqref{JCm_1} together with \eqref{meandy_jump}, we get the effective leading-order model as an effective time-dependent wave equation for the displacement $u_0$: 
\begin{equation}
\partial_t \left(\alpha_0(t)\,\partial_t \overline{u}_0\right)-\beta_0(t)\,\partial^2_{xx} \overline{u}_0=f,
\label{WaveAdim0}
\end{equation}
where we have introduced the following effective parameters
\begin{equation}
\label{def_effective_alpha_beta_gamma}
\alpha_0(t)=\mean{\alpha}+\sum_{\ell=1}^{N}\mathscr{m}_\ell(t),\qquad \beta_0(t)=\mean{\beta(1+\partial_yP_1(\cdot,t))}. 
\end{equation}
The effective mass density $\alpha_0(t)$ is defined explicitly in terms of the physical parameters. One can look for similar expressions for $\beta_0(t)$ . 
Integration of the PDE in \eqref{pb_cell_P1} on $(y_{\ell-1},y_{\ell})$ gives
\begin{equation}
\beta\left(1+\partial_y P_1\right)=\mathscr{C}_{\partial P_1,\ell}(t),\qquad \ell=2,\cdots,N,
\end{equation}
for some unknown functions $\mathscr{C}_{\partial P_1,\ell}$. The continuity of flux in \eqref{pb_cell_P1} ensures that $\mathscr{C}_{\partial P_1,\ell}$ does not depend on $\ell$, and hence 
\begin{equation}
\mathscr{C}_{\partial P_1,\ell}(t) := \beta_0(t)= \beta\left(1+\partial_y P_1\right).
\label{Beta0Def}
\end{equation}
It follows that
\begin{equation}
\partial_y P_1=\frac{\beta_0}{\beta}-1.
\label{dyP1}
\end{equation}
Averaging \eqref{dyP1} and using \eqref{meandy_jump} yields
\begin{equation}
-\sum_{\ell=1}^{N}\jump{P_1}_\ell=\beta_0\mean{\frac{1}{\beta}}-1.
\end{equation}
Finally, the jump condition satisfied by $P_1$ in \eqref{pb_cell_P1} gives
 the non-dimensionalised time-dependent leading-order effective rigidity
\begin{equation}
\beta_0(t)=\left(\mean{\frac{1}{\beta}}+\sum_{\ell=1}^{N}\mathscr{c}_\ell(t)\right)^{-1} > 0.
\label{Beta0}
\end{equation}
\vspace{-10pt}
\paragraph{\bf Dimensionalised coordinates.}From  \eqref{KMadim}, \eqref{Adim} and \eqref{AlphaBeta}, we obtain the dimensionalised leading-order model
\begin{equation}
\label{leading_order_dim}
\partial_T\left[\rho_0(T)\partial_T\overline{U}_0(X,T)\right]=E_0(T)\,\partial^2_{XX}\overline{U}_0(X,T)+F(X,T),
\end{equation}
together with the dimensionalised effective density and effective rigidity: 
\begin{equation}
\rho_0(T)=\mean{\rho}+\frac{1}{h}\sum_{\ell=1}^{N}\mathscr{M}_\ell(T),\qquad \frac{1}{E_0(T)}=\mean{\frac{1}{E}}+\frac{1}{h}\sum_{\ell=1}^{N} \mathscr{C}_\ell(T).
\label{E0R0}
\end{equation}
Two limit cases are interesting:
\begin{itemize}
\vspace{-10pt}
\item in the case of perfect contact ($\mathscr{M}_\ell=0$, $\mathscr{C}_\ell=0$)
, then one recovers the usual effective parameters: $\rho_0=\mean{\rho}$ and $1/E_0=\mean{1/E}$;
\item in the case of one non-modulated imperfect interface per cell ($N=1$), then one recovers the effective parameters given in \cite{bellisJotMaPoS2021}: $\rho_0=\mean{\rho}+\frac{\mathscr{M}_1}{h}$ and $\frac{1}{E_0}=\mean{\frac{1}{E}}+\frac{\mathscr{C}_1}{h}$.
\end{itemize}

\subsection{Properties}
In this section, we analyse some properties of the obtained leading-order effective model. The first important property is that, by solely modulating \textit{interface properties} one gets an effective model whose effective \textit{bulk properties} are time-dependent. This network of modulated interfaces can therefore be seen as a way to create time-varying metamaterials which are known to exhibit unusual phenomena such as time reflection and time refraction, unusual topology \cite{Lustig2018}, frequency conversion \cite{Salehi2022}, as well as unidirectional and parametric amplification \cite{kiorpelidisPRB2024,kimPRE2023,TorrentPRB2018}. We focus here on some of these unusual phenomena.

\subsubsection{$k$-gaps for a modulation periodic in time}\label{Sec:Prop_kgaps}
Let us consider the case of a periodic time modulation of the interface parameters with period $\tau$. Consequently, the leading-order effective parameters \eqref{E0R0} in the leading-order effective model \eqref{leading_order_dim} are also $\tau$-periodic in time. Using the space-domain Fourier transform,  \eqref{leading_order_dim} then becomes:
\begin{equation}
   \partial_T\left(\rho_0(T)\,\partial_T \hat{U}(T)\right)=-k^2E_0(T)\,\hat{U}(T),
   \label{eq:Wave1DFt}
\end{equation}
where $\hat{U}$ refers to the Fourier transform in space of $\overline{U}_0$, and $k$ is the wavenumber. The Floquet theorem ensures that the displacement field is the product of a $\tau$-periodic function and a phase shift. Then one can write 
\begin{equation}
  \hat{U}(T)=U^\sharp(T)\,e^{i\omega T},
  \label{eq:UBlocht}
\end{equation}
with $U^\sharp(T+\tau)=U^\sharp(T)$ and $\omega$ is the Bloch angular frequency evolving in the first Brillouin zone, \textit{i.e} $\omega\in[0,\frac{\pi}{\tau}]$.
For piecewise constant parameters $\rho_0$ and $E_0$, the dispersion relation is easily found using \eqref{eq:Wave1DFt} and \eqref{eq:UBlocht} using e.g.\ the transfer matrix method. For smooth modulations, the derivation of the dispersion relation relies on plane wave expansions and is detailed in Appendix~\ref{AppPWE}.  

In both modulations, gaps in wavenumber, i.e.\ $k$-gaps may occur. They are the time-counterpart of the usual gaps in frequency observed for materials periodic in space and constant in time. It is non-usual that this kind of behaviour, associated to a high dispersion, is encapsulated by a leading-order model. In our case, it is due to the fact that we performed homogenisation assuming a small parameter in space, but without any further assumption on the time. Moreover, contrary to a gap in frequency where the fields are evanescent, in a $k$-gap one gets amplification of the solution. We will see later on that this amplification is well described by the leading-order model.
\vspace{-5pt}
\subsubsection{Reciprocity}\label{Sec:Reciprocity}
Regardless of the choice of modulation, \eqref{leading_order_dim} is a wave equation with time-dependent parameters. It remains invariant by changing $x$ into $-x$, and the effective medium is therefore reciprocal. 

\vspace{-5pt}
\section{High-order homogenisation}\label{Sec:HomogOrdre2}

To take into account dispersive effects that increase drastically as the modulation and source frequency increase, one needs higher-order homogenised models. Furthermore, by analogy with wave-like bulk modulations \cite{Touboul2024,Touboul2024b},  one might wonder whether these higher-order models are able to describe the non-reciprocal effects missed by the leading-order homogenisation. This section is dedicated to deriving such higher-order effective models. 
\subsection{First-order effective model}
Injecting the displacements $u_0$ \eqref{macro_v0_sigma_0} and $u_1$ \eqref{form_u1} into \eqref{JC_o2} and using the leading-order wave equation \eqref{WaveAdim0} allows one to write $u_2$ as follows: 
\begin{equation}
\begin{aligned}
u_2(x,y,t)&=\overline{u}_2(x,t)+P_1(y,t)\,\partial_x\overline{u}_1(x,t)\\
&+P_2(y,t)\,\partial^2_{xx}\overline{u}_0(x,t)+P_3(y,t)\,\partial^2_{tt} \overline{u}_0(x,t)+P_4(y,t)\,\partial_t \overline{u}_0(x,t),
\end{aligned}
\label{U2}
\end{equation}
where the correctors $P_{2,3,4}$ are solutions of the cell problems $(\mathscr{P} _{2,3,4})$ defined in Appendix~\ref{AppCellU2}. 

From the ${\cal O}(\eta^{1})$ term in the wave equation \eqref{EDPscaled}, the ${\cal O}(\eta^{2})$ term in the jump of stress \eqref{CascadeJCS}, and the ${\cal O}(\eta^{3})$ term in the jump of displacement \eqref{CascadeJCu}, one obtains
\begin{subnumcases}{\label{EDP-eta1}}
\ds \alpha\,\partial^2_{tt}\,u_1=\partial_y\left(\beta\left(\partial_x u_2+\partial_y u_3\right)\right)+\partial_x\left(\beta\left(\partial_x u_1+\partial_y u_2\right)\right), \label{EDP-eta1a}\\ [-6pt]
\ds \jump{u_3}_\ell=\mathscr{c}_\ell(t)\,\dmean{\beta\,(\partial_x u_2+\partial_y u_3)}_\ell,\label{EDP-eta1b}\\ [-6pt]
\ds \jump{\beta\,(\partial_x u_2+\partial_y u_3)}_\ell=\partial_t\left(\mathscr{m}_\ell(t)\,\dmean{\partial_t u_1}_{\ell}\right),\label{EDP-eta1c}
\end{subnumcases}
for $\ell=1,\cdots,N$. Averaging \eqref{EDP-eta1a} and using \eqref{EDP-eta1c} leads to
\begin{equation}
\mean{\alpha\,\partial^2_{tt}u_1}=-\sum_{\ell=1}^{N}\partial_t\left(\mathscr{m}_\ell(t)\,\dmean{\partial_t u_1}_{\ell}\right)+\mean{\beta\,(\partial^2_{xx}u_1+\partial^2_{xy}u_2)}.
\end{equation}
From the displacements \eqref{macro_v0_sigma_0}, \eqref{form_u1} and \eqref{U2}, one obtains
\begin{align}
\ds \partial_t\left(\alpha_0(t)\,\partial_t\overline{u}_1\right)&-\beta_0(t)\,\partial^2_{xx}\overline{u}_1=\mean{\beta\left(P_1+\partial_y P_2\right)}\,\partial^3_{xxx}\overline{u}_0-\mean{\alpha\,\partial^2_{tt}\left(P_1\,\partial_x\overline{u}_0\right)} \nonumber\\[-6pt]
\ds &+\mean{\beta\,\partial_y P_3}\partial^3_{ttx}\overline{u}_0+\mean{\beta\,\partial_y P_4}\partial^2_{tx}\overline{u}_0-\sum_{\ell=1}^{N}\partial_t\left(\mathscr{m}_\ell(t)\,\dmean{\partial_t P_1}_{\ell}\,\partial_x\overline{u}_0\right).
\end{align}
Expansions of the time derivatives lead to
\begin{equation}
\begin{array}{l}
\ds \partial_t\left(\alpha_0(t)\,\partial_t\overline{u}_1\right)-\beta_0\,\partial^2_{xx}\,\overline{u}_1=\mean{\beta\left(P_1+\partial_y P_2\right)}\partial^3_{xxx}\overline{u}_0\\ [0pt]
\hspace{1cm} + \ds\left(\mean{-\alpha\,\partial^2_{tt}P_1}-\sum_{\ell=1}^{N}\left(\mathscr{m}'_\ell(t)\, \partial_t \dmean{P_1}+\mathscr{m}_\ell(t)\,\partial^2_{tt}\dmean{P_1}\right)\right)\partial_x\overline{u}_0 \\ [0pt]
\hspace{1cm} + \ds \left(\mean{-2\,\alpha\,\partial_t P_1}+\mean{\beta\,\partial_y P_4}-\sum_{\ell=1}^{N}\left(\mathscr{m}'_\ell(t)\,\dmean{P_1}+2\,\mathscr{m}_\ell(t)\,\partial_t \dmean{P_1}\right)\right)\partial_{xt}^2\overline{u}_0 \\ [0pt]
\hspace{1cm} + \ds\left(\mean{-\alpha\,P_1}+\mean{\beta\,\partial_y P_3}-\sum_{\ell=1}^{N}\left(\mathscr{m}_\ell(t)\,\dmean{P_1}\right)\right) \partial^3_{xtt}\overline{u}_0.
\end{array}
\end{equation}
From \eqref{P1dP2} and the variational identities \eqref{IRiii}--\eqref{IRvii}, the right-hand-side of the latter equation vanishes. Finally, the first-order term $\overline{u}_1$ satisfies the following time-dependent wave equation
\begin{equation}
\partial_t \left(\alpha_0(t)\,\partial_t \overline{u}_1\right)-\beta_0(t)\,\partial^2_{xx} \overline{u}_1=0.
\label{WaveAdim1}
\end{equation}
The first-order effective wave equation has the same structure as the leading-order one \eqref{WaveAdim0}, but without a source term $f$. Therefore, {the effective equation is unchanged at the first order. In particular, if we choose zero initial conditions the first-order macroscopic field remains identically equal to zero}.


\subsection{Second-order effective model}

\subsubsection{Derivation}

Injecting the displacements $u_0$ \eqref{macro_v0_sigma_0}, $u_1$ \eqref{form_u1} and $u_2$ \eqref{U2} into the PDE \eqref{EDP-eta1a} leads to
\begin{align}
    \ds \alpha\,\partial^2_{tt}\overline{u}_1 &+\alpha\,\partial^2_{tt}P_1\,\partial_x \overline{u}_0+\alpha\,P_1\,\partial^3_{ttx}\overline{u}_0+2\,\alpha\,\partial_tP_1\,\partial^2_{tx}\overline{u}_0=\beta\,\partial^2_{xx}\overline{u}_1+\beta\,P_1\,\partial^3_{xxx}\overline{u}_0+\beta\,\partial_y P_1\,\partial^2_{xx}\overline{u}_1 \nonumber\\[0pt]
 \ds &+\beta\,\partial_y P_2\,\partial^3_{xxx}\overline{u}_0+\beta\,\partial_y P_3\,\partial^3_{ttx}\overline{u}_0+\beta\,\partial_y P_4\,\partial^2_{tx}\,\overline{u}_0+\partial_y\beta\, \partial_x\overline{u}_2+\partial_y(\beta\,P_1)\,\partial^2_{xx}\overline{u}_1 \nonumber\\[0pt]
\ds &+\partial_y(\beta\,P_2)\,\partial^3_{xxx}\overline{u}_0+\partial_y(\beta\,P_3)\,\partial^3_{ttx}\overline{u}_0+\partial_y(\beta\,P_4)\,\partial^2_{tx}\overline{u}_0+\partial_y(\beta\,\partial_y u_3).
\label{Ordre2-A}
\end{align}
It follows that
\begin{align}
    \ds u_3(x,y,t)&=\overline{u}_3(x,t)+P_1(y,t)\,\partial_x\overline{u}_2(x,t)+P_2(y,t)\,\partial^2_{xx}\overline{u}_1(x,t)+P_3(y,t)\,\partial^2_{tt} \overline{u}_1(x,t) \nonumber\\ 
&+P_4(y,t)\,\partial_t \overline{u}_1(x,t)+P_5(y,t)\,\partial^3_{xxx}\overline{u}_0(x,t)+P_6(y,t)\,\partial^3_{ttx}\overline{u}_0(x,t) \nonumber\\ 
&+P_7(y,t)\,\partial^2_{tx}\overline{u}_0(x,t)+P_8(y,t)\,\partial_x\overline{u}_0(x,t),
\label{U3}
\end{align}
where the correctors $P_{5,6,7,8}$ are solutions of the cell problems $(\mathscr{P}_{5,6,7,8})$ given in Appendix \ref{AppCellU3}. From the ${\cal O}(\eta^{2})$ term in the wave equation \eqref{EDPscaled}, the ${\cal O}(\eta^{3})$ term in the jump of stress \eqref{CascadeJCS}, and the ${\cal O}(\eta^{4})$ term in the jump of displacement \eqref{CascadeJCu}, one obtains
\begin{subnumcases}{\label{EDP-eta2}}
\ds \alpha\,\partial^2_{tt}\,u_2=\partial_y\left(\beta\left(\partial_x u_3+\partial_y u_4\right)\right)+\partial_x\left(\beta\left(\partial_x u_2+\partial_y u_3\right)\right), \label{EDP-eta2a}\\ [-6pt]
\ds \jump{u_4}_\ell=\mathscr{c}_\ell(t)\,\dmean{\beta\,(\partial_x u_3+\partial_y u_4)}_\ell,\label{EDP-eta2b}\\ [-6pt]
\ds \jump{\beta\,(\partial_x u_3+\partial_y u_4)}_\ell=\partial_t\left(\mathscr{m}_\ell(t)\,\dmean{\partial_t u_2}_{\ell}\right).\label{EDP-eta2c}
\end{subnumcases}
Averaging \eqref{EDP-eta2a} and using \eqref{EDP-eta2c} leads to
\begin{equation}
\mean{\alpha\,\partial^2_{tt}u_2}=-\sum_{\ell=1}^{N}\partial_t\left(\mathscr{m}_\ell(t)\,\dmean{\partial_t u_2}_{\ell}\right)+\mean{\beta\,(\partial^2_{xx}u_2+\partial^2_{xy}u_3)}.
\end{equation}
The displacements \eqref{macro_v0_sigma_0}, \eqref{form_u1}, \eqref{U2} and \eqref{U3} can now be injected into \eqref{EDP-eta2a}. Using \eqref{P1dP2}, \eqref{P2dP5} and the variational identities \eqref{IRiii}--\eqref{IRvii}, many terms vanish in the process. Finally, the second-order term $\overline{u}_2$ can be shown to satisfy the following effective equation
\begin{equation}
\partial_t \left(\alpha_0(t)\,\partial_t \overline{u}_2\right)-\beta_0(t)\,\partial^2_{xx} \overline{u}_2={\cal F}(\overline{u}_0),
\label{WaveAdim2}
\end{equation}
with forcing $\mathcal{F}$ depending on leading-order terms and given by
\begin{align}
{\cal F}(\overline{u}_0) &= \ds \mathfrak{A}\,\partial^4_{xxtt}\overline{u}_0 + \mathfrak{B}\,\partial^4_{tttt}\overline{u}_0 + \ds \mathfrak{C}\,\partial^3_{txx}\overline{u}_0 + \ds \mathfrak{D}\,\partial^3_{ttt}\overline{u}_0 + \ds \mathfrak{E}\,\partial^2_{xx}\overline{u}_0+ \ds \mathfrak{F}\,\partial^2_{tt}\overline{u}_0 + \ds \mathfrak{G}\,\partial_{t}\overline{u}_0,
\label{FU0}
\end{align}
with
\begin{align*}
\mathfrak{A}&= \ds \mean{\beta(P_3+\partial_y P_6)-\alpha\,P_2}-\sum_{\ell=1}^{N}\mathscr{m}_\ell(t)\,\dmean{P_2}_\ell;  \nonumber\\[-5pt]   \mathfrak{B}&=\mean{-\alpha\,P_3}-\sum_{\ell=1}^{N}\mathscr{m}_\ell(t)\dmean{P_3}_\ell; \nonumber\\[-5pt]  
\mathfrak{C}&= \ds \mean{\beta(P_4+\partial_y P_7)-2\,\alpha\partial_t P_2}-\sum_{\ell=1}^{N}\mathscr{m}'_\ell(t)\dmean{P_2}_\ell-\sum_{\ell=1}^{N}\mathscr{m}_\ell(t)\,2\dmean{\partial_t P_2}_\ell; \nonumber \\[-5pt]  
\mathfrak{D}&= \ds \mean{-\alpha\,P_4-2\,\alpha\,\partial_t P_3}-\sum_{\ell=1}^{N}\mathscr{m}'_\ell(t)\,\dmean{P_3}_\ell-\sum_{\ell=1}^{N}\mathscr{m}_\ell(t)\,\dmean{2\,\partial_tP_3+P_4}_\ell; \nonumber \\[-5pt] 
\mathfrak{E}&= \ds \mean{-\alpha\,\partial^2_{tt}P_2+\beta\,\partial_y P_8}-\sum_{\ell=1}^{N}\mathscr{m}'_\ell(t)\,\dmean{\partial_t P_2}_\ell-\sum_{\ell=1}^{N}\mathscr{m}_\ell(t)\,\dmean{\partial^2_{tt} P_2}_\ell; \nonumber \\[-5pt] 
\mathfrak{F}&= \ds \mean{-\alpha\,\partial^2_{tt}P_3-2\,\alpha\,\partial_t P_4}-\sum_{\ell=1}^{N}\mathscr{m}'_\ell(t)\,\dmean{\partial_t P_3+P_4}_\ell-\sum_{\ell=1}^{N}\mathscr{m}_\ell(t)\,\dmean{\partial^2_{tt}P_3+2\,\partial_t P_4}_\ell; \nonumber \\[-5pt] 
\mathfrak{G}&= \ds \mean{-\alpha\,\partial^2_{tt}\,P_4}-\sum_{\ell=1}^{N}\mathscr{m}'_\ell(t)\,\dmean{\partial_t P_4}_\ell-\sum_{\ell=1}^{N}\mathscr{m}_\ell(t)\,\dmean{\partial^2_{tt} P_4}_\ell.
\label{FU0}
\end{align*}

\begin{Remark}
The forcing ${\cal F}(\overline{u}_0)$ does not cancel out, even in the limiting case of perfect contact. 
Indeed for perfect contact, all the cell problems solutions depend on $y$ only, and we also have  $P_4=P_7=P_8=0$, therefore ${\cal F}(\overline{u}_0)$ simply reads:
\begin{equation}
{\cal F}(\overline{u}_0) =\ds \mean{\beta(P_3+\partial_y P_6)-\alpha\,P_2}\partial^4_{xxtt}\overline{u}_0-\ds \mean{\alpha\,P_3}\partial^4_{tttt}\overline{u}_0.
\label{FU0_perf_contact}
\end{equation}
By using variational identities for the cell problems, one can show that the source term reduces to 
\begin{equation}
    {\cal F}(\overline{u}_0) =\ds \mean{\alpha P_1^2-2\alpha P_2}\partial^4_{xxtt}\overline{u}_0+\ds \mean{(\partial_y P_3)^2}\partial^4_{tttt}\overline{u}_0.
\end{equation}
In particular, if $\alpha$ or $\beta$ is constant we recover the limit cases of \cite{Touboul2024} i.e.\ $\langle P_1^2\rangle\partial^4_{xxtt}\overline{u}_0$ and $\langle(\beta\partial_y P_3)^2\rangle \partial^4_{tttt}\overline{u}_0$, respectively. 
\end{Remark}

\subsubsection{Total mean field at order 2}\label{SecHomogTotal}
From \eqref{ansatz}, the second-order total mean field $\overline{u}^{(2)}$ is defined by
\begin{equation}
\overline{u}^{(2)}(x,t)=\overline{u}_0(x,t)+\eta\,\overline{u}_1(x,t)+\eta^2\,\overline{u}_2(x,t).
\label{U2TotalDef}
\end{equation}
The effective equations at leading order \eqref{WaveAdim0}, first order \eqref{WaveAdim1} and second order \eqref{WaveAdim2} are collected. Neglecting the higher-order terms, we deduce that $\overline{u}^{(2)}$ satisfies
\begin{equation}
\partial_t \left(\alpha_0(t)\,\partial_t \overline{u}^{(2)}\right)-\beta_0(t)\,\partial^2_{xx} \overline{u}^{(2)}=f+\eta^2{\cal F}(\overline{u}^{(2)}).
\label{WaveTotal2}
\end{equation}
\begin{Proposition}[Reciprocity]
No odd-order spatial derivatives are involved in \eqref{WaveTotal2} and \eqref{FU0}. Therefore the equations remain invariant by changing $x$ into $-x$, and the effective model is reciprocal.
\label{PropRecipro}
\end{Proposition}


\subsection{Computation of the second-order effective model in a limit case}
To see how much further the second-order model pushes the limits of validity of the homogenisation process, we choose a limit case for which the effective parameters can be computed analytically. More precisely, we choose constant bulk parameters ($\alpha=\beta=1$) and $\mathscr{m}_\ell = 0$ for $\ell=1,\cdots,N$, having therefore continuity of the stress but a time-modulated jump condition for the displacement. 

\subsubsection{Computation of the effective parameters}
In this case one gets $\alpha_0=1$, $P_3=P_4=0$, and the source term in \eqref{WaveTotal2} reduces to 
\begin{equation}
\label{Simpl_F}
{\cal F}(\overline{u}^{(2)}) = \ds \mean{\partial_y P_6}\partial^4_{xxtt}\overline{u}^{(2)}+\ds \mean{\partial_y P_7}\,\partial^3_{txx}\overline{u}^{(2)}+\ds \mean{\partial_y P_8}\,\partial^2_{xx}\overline{u}^{(2)}.
\end{equation} 
In this limit case, averaging of $\eqref{CellPbP6}\times P_1-\eqref{pb_cell_P1}\times P_6$ leads to
\begin{equation}
\mean{\partial_y P_6} =\mean{P_1^2},
\end{equation}
averaging of $\eqref{CellPbP7}\times P_1-\eqref{pb_cell_P1}\times P_7$ leads to
\begin{equation}
\mean{\partial_y P_7} =2\mean{\partial_t P_1 P_1} = \partial_t \mean{P_1^2},
\end{equation}
while 
averaging of $\eqref{CellPbP8}\times P_1-\eqref{pb_cell_P1}\times P_8$ yields
\begin{equation}
\mean{\partial_y P_8} =\mean{ P_1\partial^2_{tt} P_1}.
\end{equation}
Therefore, the second-order mean field $\overline{u}^{(2)}$ satisfies
\begin{equation}
\partial_t \left(\alpha_0(t)\,\partial_t \overline{u}^{(2)}\right)-(\beta_0(t)+\eta^2{\mathfrak{b}}_1(t))\,\partial^2_{xx} \overline{u}^{(2)}-\eta^2\partial_t({\mathfrak{b}}_2(t)\,\partial^3_{txx}\overline{u}^{(2)})=f,
\label{WaveTotal2_simpl}
\end{equation}
with two additional effective parameters defined by
\begin{equation}
\label{param_limit_case}
{\mathfrak{b}}_1(t) = \mean{P_1\partial^2_{tt}  P_1},\quad {\mathfrak{b}}_2(t) = \mean{P_1^2}.
\end{equation}

\subsubsection{Effective equation in dimensionalised coordinates and energy analysis}

From \eqref{KMadim}, \eqref{Adim} and \eqref{AlphaBeta}, one obtains the dimensionalised second-order model in the limit case of constant bulk parameters and no interface inertia:
\begin{equation}
\label{eq:dim_order2}
\rho\,\partial^2_{TT}U-\left(E_0(T)+h^2\rho\,\mathcal{B}_1(T)\right)\partial^2_{XX}U-h^2\rho\,\partial_T\left(\mathcal{B}_2(T)\,\partial^3_{XXT}U\right)=F(X,T),
\end{equation}
where, under the abuse of notation $P_1(.,t)\mapsto P_1(.,T)$,
\begin{equation}
\label{param_limit_case-Dim}
{\cal B}_1(T) = \mean{P_1\partial^2_{TT}  P_1},\quad {\cal B}_2(T) = \mean{P_1^2}.
\end{equation}
The wave equation is thus enriched by a fourth-order mixed derivative in space and time. We take the product of \eqref{eq:dim_order2} with $\partial_T U$ and integrate over $\mathbb{R}$. Integration by parts leads to the following energy balance.
\begin{Proposition}
Let
\begin{equation}
{\cal E}(T)=\frac{1}{2}\int_\mathbb{R}\rho\left(\partial_T U\right)^2dX+\frac{1}{2}\int_\mathbb{R}\left(E_0(T)+h^2\rho\,\mathcal{B}_1(T)\right)\left(\partial_X U\right)^2dX+\frac{h^2}{2}\int_\mathbb{R}\rho\,\mathcal{B}_2(T)\left(\partial^2_{XT}U\right)^2dX.
\label{NRJ2}
\end{equation}
Introducing the power of external forces 
\begin{equation}
{\cal P}(t)=\int_\mathbb{R}F\,\partial_TU\,dX,
\label{Power-NRJ2}
\end{equation}
then one has
\begin{equation}
\frac{d}{dT}{\cal E}(T)={\cal P}(T)-\frac{1}{2}\int_\mathbb{R}\left(E_0'(T)+h^2\rho\,\mathcal{B}_1'(T)\right)\left(\partial_X U\right)^2dX-\frac{h^2}{2}\int_\mathbb{R}\rho\,\mathcal{B}_2'(T)\left(\partial^2_{XT}U\right)^2dX.
\label{Balance-NRJ2}
\end{equation}
\label{propNRJ2}
\end{Proposition}
\vspace*{-20pt}
\noindent Three remarks arise from Proposition \ref{propNRJ2}:
\begin{itemize}
\vspace{-10pt}
\item the parameter ${\cal B}_2$ in \eqref{param_limit_case-Dim} is non-negative.  The parameter $\mathcal{B}_1$ can be negative so that $E_0(T)+h^2\rho\,\mathcal{B}_1(T)$ can reach negative values if $h$ is large or $\mathscr{C}_\ell$ varies too quickly, which contradicts the assumptions of the asymptotic expansion. Therefore, ${\cal E}\geq 0$ is the bulk energy in the context of this paper;
\item if there are no body force $(F=0$) and no modulation (${\mathcal B}_1'={\mathcal B}_2'=0$), then the energy is conserved;
\item if $F=0$ but $\mathscr{C}_\ell$ varies with time, then the total mechanical energy may vary,
due to the power of external forces required to modify the interface stiffness; the sign of the right hand side in \eqref{Balance-NRJ2} is arbitrary, so that ${\cal E}$ may increase or decrease.
\end{itemize}


\section{Numerical simulations}\label{Sec:Num}

\subsection{Numerical methods}\label{Sec:Methods}

Different types of time-domain numerical methods are used, depending on the model studied. In each case, one introduces a uniform mesh size $\Delta X=L/N_X$ (where $L$ is the length of the domain and $N_X$ is the number of grid nodes) and a time step $\Delta T$. The latter depends on a Courant-Friedrichs-Lewy (CFL) condition of stability. Lastly, given $Z(X,T)$, one denotes $Z_m^n$ an approximation of $Z(X_m=m\,\Delta X,T_n=n\,\Delta T)$.

\vspace{-10pt}
\paragraph{Microstructured model.}The momentum equation with Hooke's law \eqref{EDPmicro} is written as a first-order velocity-stress system of PDEs. It is solved using a fourth-order finite difference scheme (ADER-4). This scheme is explicit in time and stable under the usual CFL condition $\max(c_h)\,\Delta T/\Delta X \leq 1$ where $c_h=\sqrt{E_h/\rho_h}$. The modulated jump conditions are discretised using an immersed interface method, as described in \cite{Darche2025}. After each time step, a numerical integration is performed to obtain the displacement field
\vspace{-10pt}
\paragraph{Leading-order effective model.}A second-order explicit finite-difference scheme is used to solve the scalar wave equation \eqref{leading_order_dim} and determine an approximation $U_m^n$ of $\overline{U}_0(X_m,T_n)$. Second-order centered approximations of the operators lead to the scheme
\begin{align}
   &U_m^{n+1}=U_m^n \\
   &+\frac{\rho_0(T_{n-1/2})\left(U_m^n-U_m^{n-1}\right)+\left(\frac{\Delta T}{\Delta X}\right)^2E_0(T_n)\left(U_{m+1}^n-2\,U_m^n+U_{m-1}^n\right)+F(X_{{m}},T_n)}{\rho_0(T_{n+1/2})}. \nonumber
    \label{DF-ordre0}
\end{align}
This scheme is explicit in time and second-order accurate in both space and time. It is stable under the usual CFL condition $\max(c_0)\Delta T / \Delta X\leq 1$, where $
    \max(c_0)=\ds\max_T\sqrt{E_0(T)/\rho_0(T)}$.

\begin{Remark}
With dissipation, the leading-order model involves a velocity-stress formulation given in Appendix \ref{Sec:Homog_Dissip}. The numerical modelling is described in Appendix \ref{SecNumDissip}.
\end{Remark}

\vspace{-20pt}
\paragraph{Second-order homogenised medium.}
{The PDE \eqref{eq:dim_order2} can be recast as:
\begin{equation}
\label{eq:dim_order2Num}
\partial^2_{TT}U-\left(\frac{E_0(T)}{\rho}+h^2\mathcal{B}_1(T)\right)\partial^2_{XX}U-h^2\mathcal{B}_2(T)\partial^4_{XXTT}U-h^2\mathcal{B}'_2(T)\partial^3_{XXT}U=\frac{1}{\rho}F(X,T).
\end{equation}}
To solve \eqref{eq:dim_order2Num}, the following quantities are introduced:
\begin{equation}
   A(T)=\sqrt{\frac{E_0(T)}{\rho}+h^2\mathcal{B}_1(T)}\frac{\Delta T}{\Delta X}, \qquad B(T) =\frac{h^2\mathcal{B}_2(T)}{\Delta X^2}, \qquad C(T)=\frac{h^2\mathcal{B}'_2(T)\Delta T}{\Delta X^2}.
\end{equation}
Then one obtains the following discretised equation :
\begin{equation}
\begin{aligned}
    \left(1+2B(T_n)+C(T_n)\right)\,U^{n+1}_m-\left(B(T_n)+\frac{C(T_n)}{2}\right)\,\left(U^{n+1}_{m+1}+U^{n+1}_{m-1}\right)&\\
    -2\,\left(1-(A(T_n))^2+2B(T_n)\right)\,U^{n}_m-\left((A(T_n))^2-2B(T_n)\right)\left(U^{n}_{m+1}+U^{n}_{m-1}\right)&\\
    +\left(1+2B(T_n)-C(T_n)\right)\,U^{n-1}_m-\left(B(T_n)-\frac{C(T_n)}{2}\right)\,\left(U^{n-1}_{m+1}+U^{n-1}_{m-1}\right)=\frac{\Delta T^2}{\rho}F(X_{{m}},T_n).
\end{aligned}
\label{DF2-ordre2}
\end{equation}
Setting $P(T)=(1+2B(T)+C(T))$ and $Q(T)=-(B(T)+\frac{C(T)}{2})$, and considering Dirichlet boundary conditions ($U_0^n=U_{N_X}^n=0$), the implicit equations \eqref{DF2-ordre2} can be rewritten as a tridiagonal system:
\begin{equation}
\text{tridiag}(Q(T_n),P(T_n),Q(T_n)) \,\textbf{U}^{n+1}=\textbf{D}(\textbf{U}^{n},\textbf{U}^{n-1}) + \frac{\Delta T^2}{\rho}\,\textbf{F}(T_n).
    \label{DF2-ordre2-system}
\end{equation}
In \eqref{DF2-ordre2-system}, $\textbf{U}^{n}=(U_1^n,\cdots,U_{N_X-1}^n)^\top$, $\textbf{D}$ combines the effects of previous time steps, and $\textbf{F}(T_n)$ is the vector of source terms. The system \eqref{DF2-ordre2-system} can be solved efficiently using Thomas' algorithm~\cite{Strickwerda2004}. A Von-Neumann analysis with static coefficients yields a CFL condition of stability, which is extended to dynamic coefficients: $\max_T A(T)\,{\Delta T}/{\Delta X}\leq 1$. 


\subsection{Numerical set-up}
\paragraph{\bf Geometry, materials and interfaces.}We consider a medium of length $L = 1000$~m, discretised on $N_X = 4000$ grid nodes. The bulk parameters are constant and correspond to those of a bar of Poly(methyl methacrylate) a.k.a.\ Plexiglas: $\rho= 1200$~kg/m$^3$ and $c=\sqrt{E/\rho} =2800$~m/s. 

The interfaces are distributed periodically with $h=10$~m. Except in Figure \ref{fig:leading_order_DD}(f), the dissipation terms are neglected:
$\mathscr{Q}^C_{\ell}=\mathscr{Q}^M_{\ell}=0$.  We choose a sinusoidal modulation for the other interface parameters according to: 
\begin{equation}
 \mathscr{C}_\ell(T)=\overline{\mathscr{C}}_\ell\left(1+\varepsilon_{C,\ell}\,\sin(\Omega_{C,\ell} T+\Phi_{C,\ell})\right),\quad
\mathscr{M}_\ell(T)=\overline{\mathscr{M}}_\ell\left(1+\varepsilon_{M,\ell}\,\sin(\Omega_{M,\ell} T+\Phi_{M,\ell})\right).
\label{KM-T}
\end{equation}

\paragraph{\bf Forcing.}The forcing term $F(X,T)$ in \eqref{EDPmicro-a} is a source point. It is defined by $F(X,T)=S(T)\,\delta(X-X_s)$ with:
\begin{equation}
S(T)=
 \sum_{m=1}^{4}a_m\sin(b_m\,\omega_c\,T) \quad \text{ if } 0<T <1/f_c, \qquad
 0  \text{ otherwise,}
\label{Source}
\end{equation}
where $\omega_c=2\pi f_c$. In the following numerical simulations, the source is located in the middle of the medium, i.e.\ at $X_s=500$~m. 


\subsection{Leading-order effective model}
\begin{figure}
\begin{center}
\begin{tabular}{ccc}
\hspace{-0.3cm}
 \includegraphics[width=0.375\linewidth]{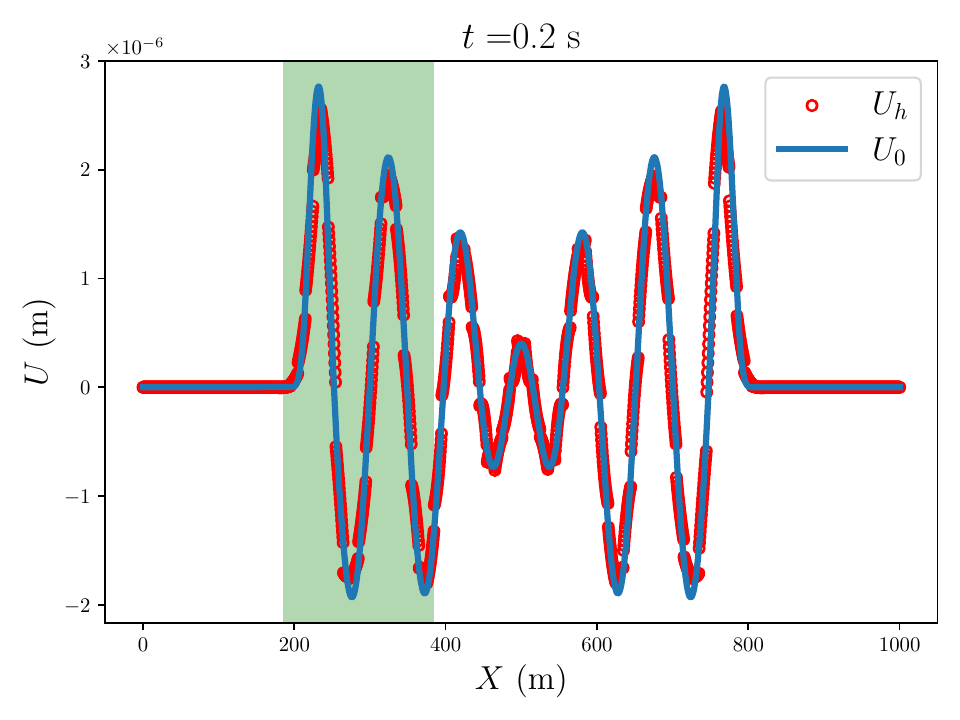} & 
\hspace{-0.4cm}
 \includegraphics[width=0.375\linewidth]{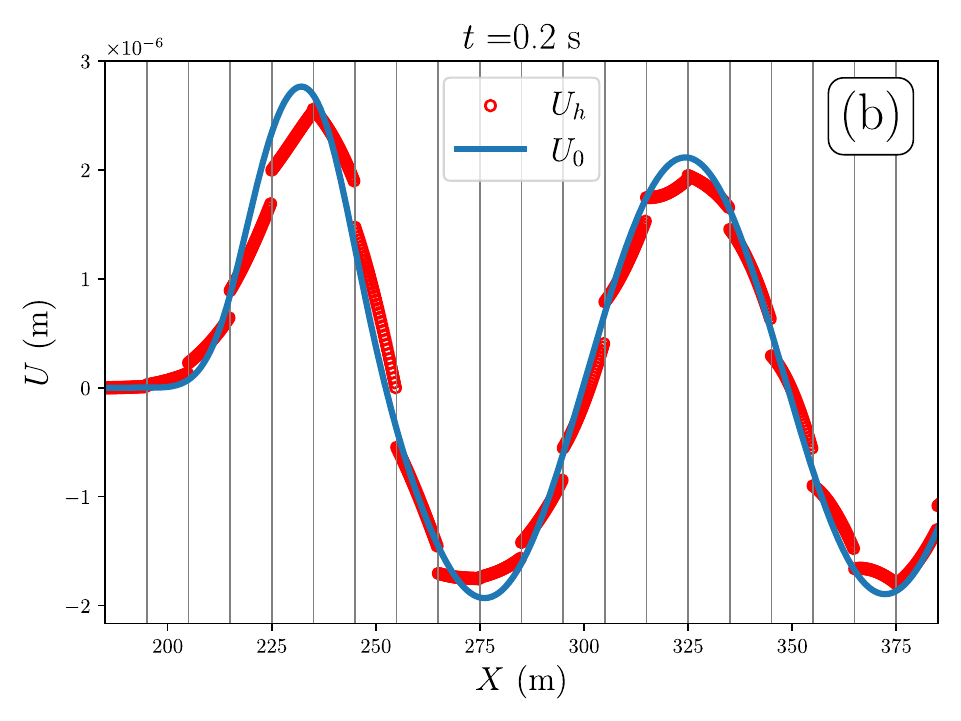}& 
\hspace{-0.4cm}
 \includegraphics[width=0.25\linewidth]{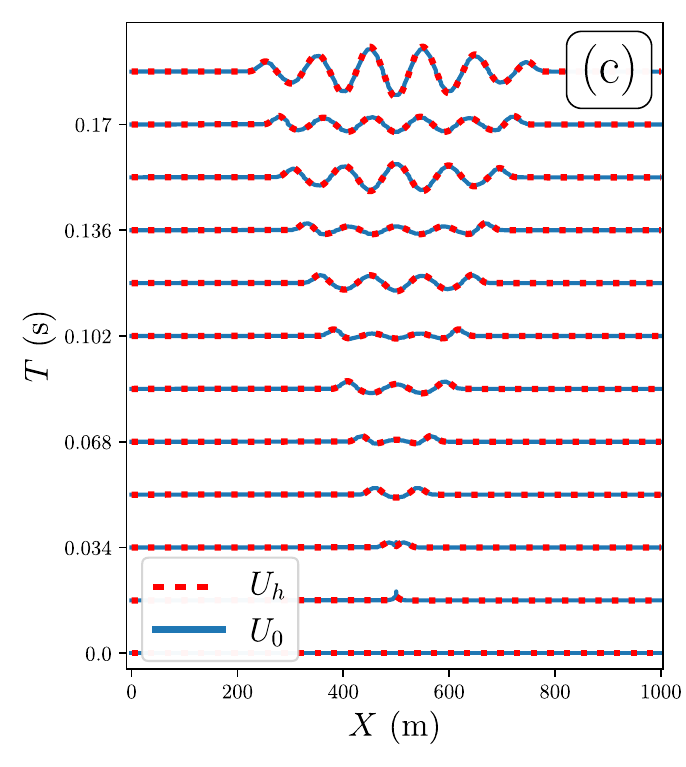} \\
\end{tabular}
\begin{tabular}{ccc}
\hspace{-0.3cm}
\includegraphics[trim={0.4cm 0.25cm 1.25cm 0.25cm},clip,width=0.34\linewidth]{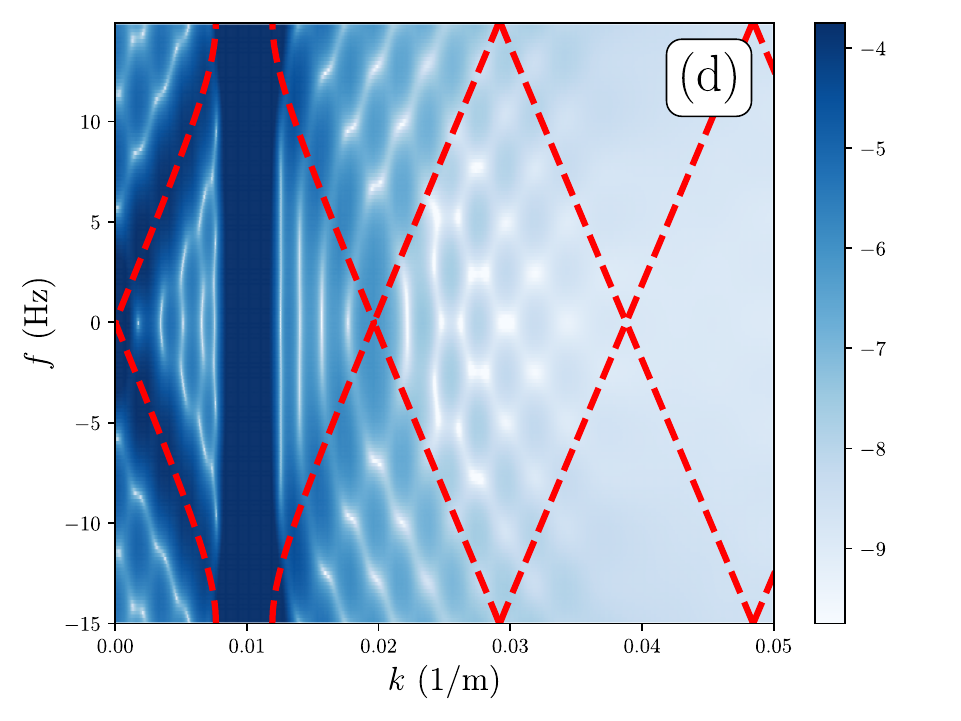}& 
\hspace{-0.3cm}
    \includegraphics[trim={0.4cm 0.25cm 2.85cm 0.25cm},clip,width=0.3\linewidth]{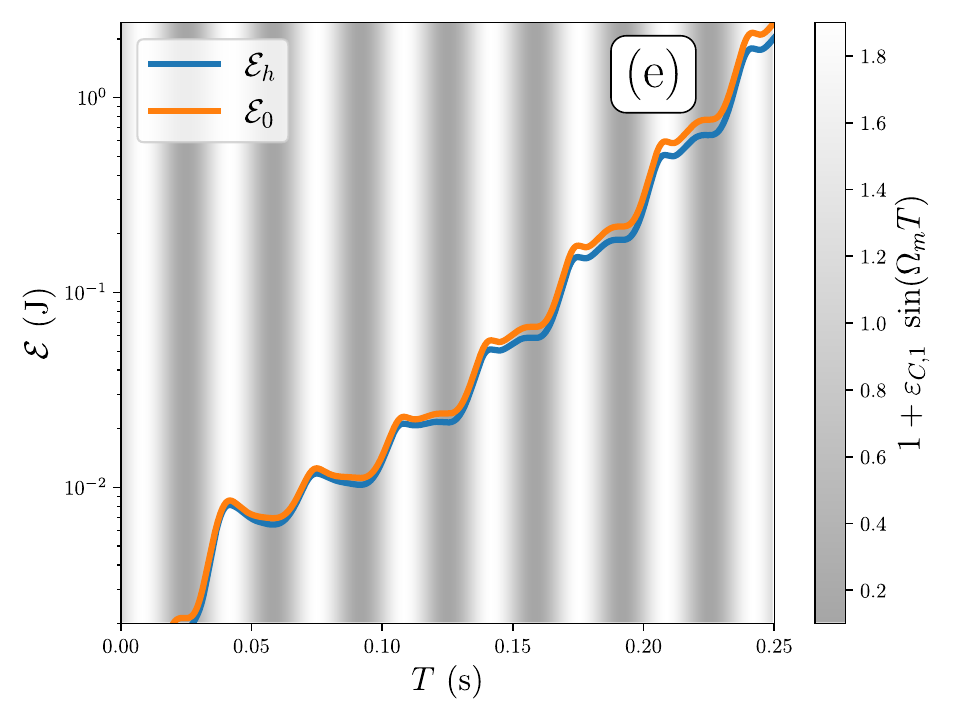} & 
    \hspace{-0.3cm}
\includegraphics[trim={0.4cm 0.25cm 0.5cm 0.25cm},clip,width=0.35\linewidth]{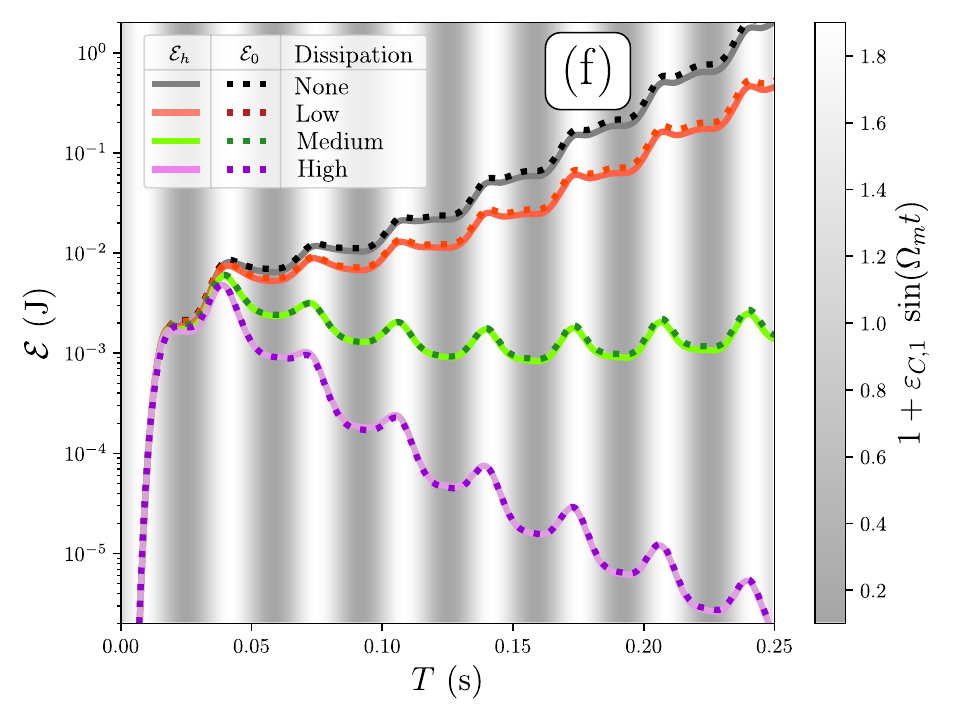}    
    \\
\end{tabular}
\end{center}
\vspace{-0.5cm}
    \caption{Case of a single modulated interface per unit cell, with $f_c=20$~Hz with $f_m=30$~Hz ($\eta_0=0.86$ and $\eta_1=2.16$). (a): Comparison of the field $U_h$ (red dots) obtained by full-field simulation in the microstructure and the leading-order field $U_0$ (blue plain line). (b): Zoom on the light-green zone of the left panel. {(c) Space-time evolution of the displacement fields: microstructured medium in red and homogenised medium in blue, showing Floquet amplification.} (d): Comparison of the dispersion diagrams. The background map is the logarithm of the norm of the double Fourier transform of the field obtained by a time-domain full-field simulation. The darkest parts therefore represent the branches of the exact dispersion relation and the location of the $k$-gaps where fields are amplified. The red dotted lines stand for the dispersion diagram of the leading-order homogenised model obtained by PWE. (e): Evolution of the energy over time for the exact field (blue) and its leading-order counterpart (orange). The background represents the time-modulation of the interface properties. (f): Time evolution of the energy when dissipative interface parameters \eqref{Q-T} are considered, illustrating the competition between intrinsic losses and Floquet amplification.} 
    \label{fig:leading_order_DD}
\end{figure}

\paragraph{Validation of the leading-order effective model.}To start with, we focus for the sake of simplicity on the case of $N=1$ modulated interface per unit cell. The mean values of the modulations are chosen such that $\overline{\mathscr{M}}_1=2\times 10^4$~kg/m$^{2}$ and $\overline{\mathscr{K}}_1=1/\overline{\mathscr{C}}_1=2.45$~GPa/m, while their amplitudes are given by $\varepsilon_{C,1}=0.9$ and $\varepsilon_{M,1}=-0.9$. With one interface, the shifts $\Phi_{C,1}$ and $\Phi_{M,1}$ have no interest, and we take them equal to zero. Both parameters are modulated with the same modulation angular frequency $\Omega_m=\Omega_{M,1}=\Omega_{C,1}$, and we introduce the frequency of modulation $f_m=\Omega_m/(2\pi)$. 
{\begin{Remark}
In the present case of a unit cell made of one single interface modulated sinusoidally in a homogeneous medium, one can rewrite the effective properties in \eqref{E0R0} as 
\begin{equation}
    \label{E0R0_impedance}
    \rho_0(T) = \overline{\rho}(1+\varepsilon_\rho \sin(\Omega_m T))\quad \text{ and } \quad \frac{1}{E_0(T) }= \frac{1}{\overline{E}}(1+\varepsilon_{E}\sin(\Omega_m T))
\end{equation}
with 
\begin{equation}
\overline{\rho}=\rho+\frac{\overline{\mathscr{M}}_1}{h},\qquad \overline{E}=\frac{Eh}{\overline{\mathscr{C}}_1E+h}, \qquad \varepsilon_\rho=\frac{\overline{\mathscr{M}}_1\varepsilon_{M,1}}{\overline{\mathscr{M}}_1+\rho h}  ,\qquad \varepsilon_E=\frac{\overline{\mathscr{C}}_1E \varepsilon_{C,1}}{\overline{\mathscr{C}}_1E+ h}.
\label{EReff}
\end{equation}
\end{Remark}
}
As shown in various works \cite{mallejacPRA2023,Koukouraki2025}, Floquet harmonics at frequencies $\omega_n=\omega_c+n\Omega_m$ are generated at each interface. In order to quantify the generation of harmonics, small parameters $\eta_n=k_n^\star\,h$ are introduced, with $k_n^\star=\frac{\omega_n}{c^\star}$ with the reference sound speed $c^\star$ chosen as 
\begin{equation}\label{Cstar}
\frac{1}{c^\star}=\sqrt{\left(\mean{\rho}+\frac{1}{h}\sum_{\ell=1}^{N}\overline{\mathscr{M}}_\ell\right)\left(\mean{\frac{1}{E}}+\frac{1}{h}\sum_{\ell=1}^{N} \overline{\mathscr{C}}_\ell\right)}.
\end{equation}
In Figure \ref{fig:leading_order_DD} (a) and (b), the leading-order homogenised field $U_0$ is compared to its counterpart $U_h$ computed by full-field simulations in the microstructure. These simulations correspond to a choice of central frequency ($f_c=20$~Hz) and modulation frequency ($f_m=30$~Hz) such that the associated so-called {small} parameters are $\eta_0=0.86$ and $\eta_1=2.16$. One can see a very good agreement despite the already high values of $\eta_0$ and $\eta_1$. It is also observed that the intrinsic dispersion due to the modulation of the interface parameters is well-described by the leading-order effective model, {as visible on the space-time evolution in Figure \ref{fig:leading_order_DD}(a-c)}. 
\vspace{-10pt}
\paragraph{Link with Floquet-Bloch analysis.}
Since the modulation is periodic in time, we also compare the dispersion diagrams of the leading-order effective model (Section 3.c.\ref{Sec:Prop_kgaps}) to the dispersion diagrams obtained for the microstructured medium by doing a double Fourier transform in space and in time. The results are presented in Figure \ref{fig:leading_order_DD}(d). The leading-order effective model captures well the location of the first $k$-gap {where the fields are amplified {(see Figure \ref{fig:leading_order_DD}(c))}}. One notes that the first branches are also well described by the leading-order effective model. In Figure \ref{fig:leading_order_DD}(e), we also represent the evolution of the energy over time for both the microstructured and leading-order model. As expected, the energy is increasing over time and this amplification phenomenon is described very accurately by the leading-order effective model.

\begin{figure}
    \centering
    \includegraphics[width=0.4\linewidth]{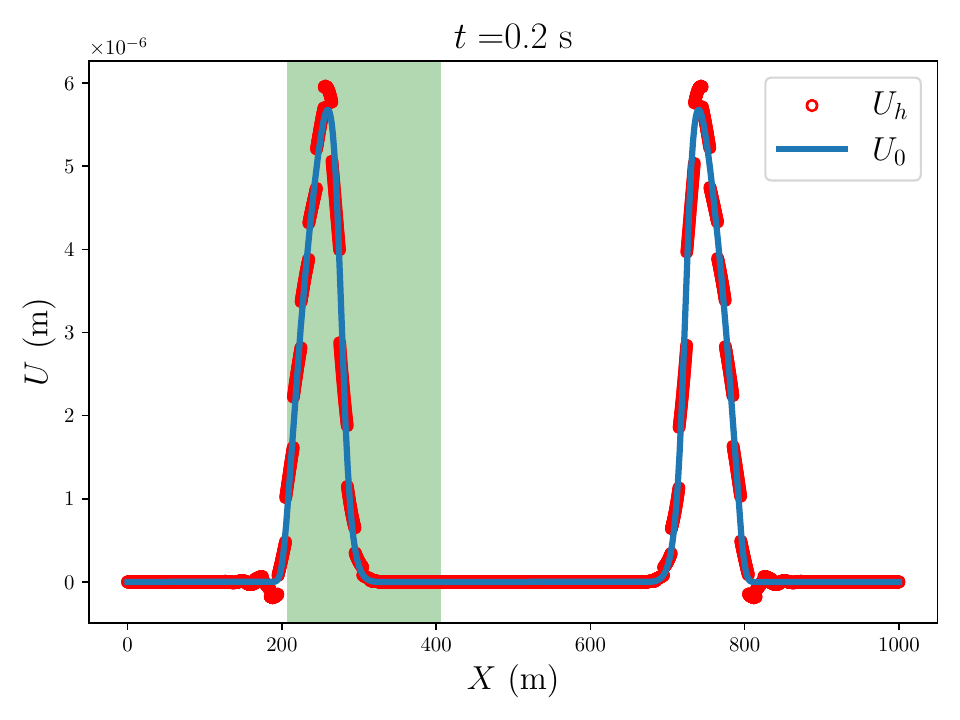}
    \includegraphics[width=0.4\linewidth]{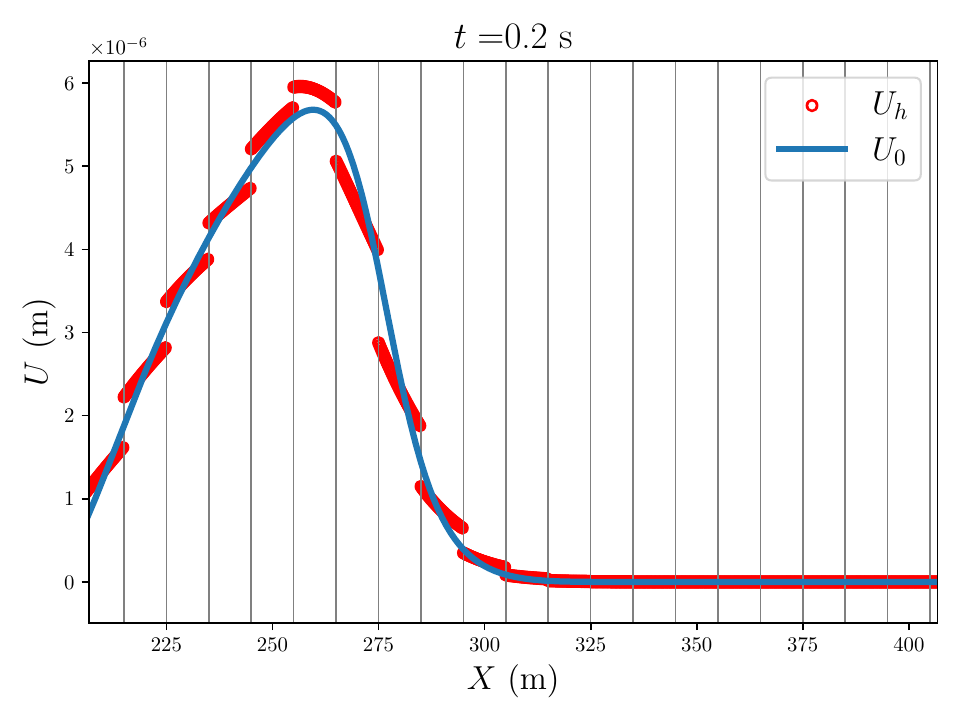} 
    \caption{Case of a single modulated interface per unit cell with interface properties set to have impedance matching \eqref{eq:impedance_matching}. (Left): comparison of the field $U_h$ (red dots) obtained by performing a full-field simulation in the microstructure and the leading-order field $U_0$ (blue plain line). (Right): Zoom on the light-green zone of the left panel. Simulations done with $f_c=10$~Hz and $f_m=20$~Hz (i.e.\ $\eta_0=0.44$ and $\eta_1=1.31$).}
    \label{fig:impedance}
\end{figure}

\begin{figure}
    \centering
    \includegraphics[width=0.4\linewidth]{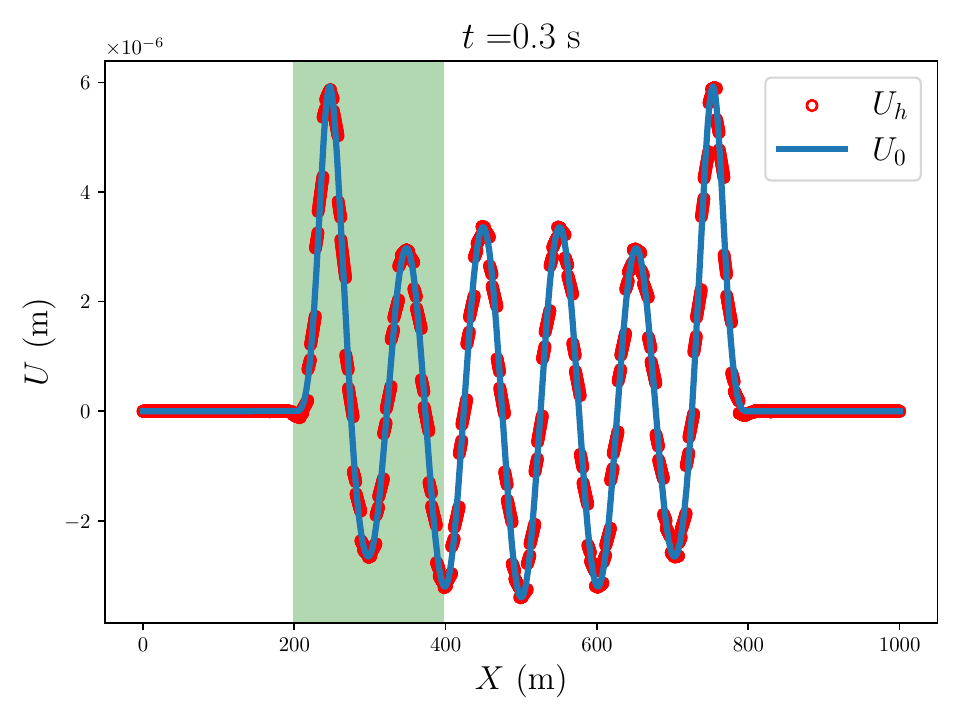}
    \includegraphics[width=0.4\linewidth]{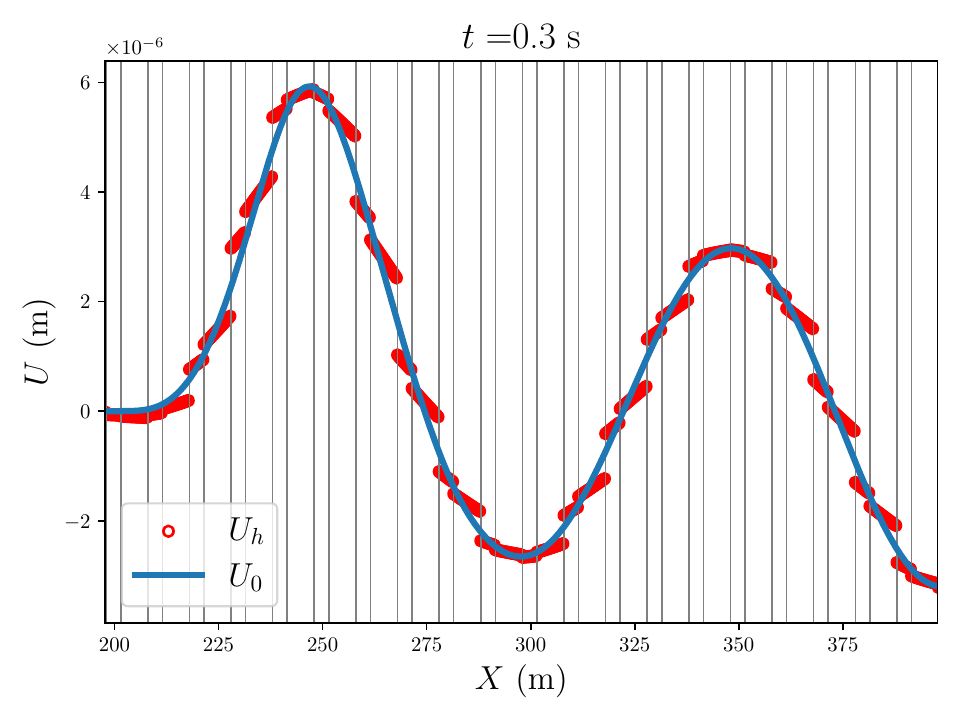} \\
    \caption{Case of two modulated interfaces per unit cell. (Left): comparison of the field $U_h$ (red dots) obtained by performing a full-field simulation in the microstructure and the leading-order field $U_0$ (blue plain line). (Right): Zoom on the light-green zone of the left panel. Simulations done with $f_c=10$~Hz and $f_m=20$~Hz  (i.e.\ $\eta_0=0.640$ and $\eta_1=1.92$). }
    \label{fig:2_interf}
\end{figure}

{Figure \ref{fig:leading_order_DD}(f) illustrates the influence of dissipation. The dissipative interface parameters are
\begin{equation}
\mathscr{Q}_{C,1}(T)=\overline{\mathscr{Q}}_{C,1}\left(1+\varepsilon_{Q_C,1}\,\sin(\Omega_m T)\right),\quad
\mathscr{Q}_{M,1}(T)=\overline{\mathscr{Q}}_{M,1}\left(1+\varepsilon_{Q_M,1}\,\sin(\Omega_m T)\right).
\label{Q-T}
\end{equation}
Three sets of values of the dissipation parameters are successively considered: $\overline{\mathscr{Q}}_{C,1}=10^{-8}$~m$^2\cdot$s/kg and $\overline{\mathscr{Q}}_{M,1}=10^{4}$~kg/(m$^2$s) for the low dissipation case,  $\overline{\mathscr{Q}}_{C,1}=5\times10^{-8}$~m$^2\cdot$s/kg and $\overline{\mathscr{Q}}_{M,1}=5\times10^{4}$~kg/(m$^2$s) for the medium one, and $\overline{\mathscr{Q}}_{C,1}=10^{-7}$~m$^2\cdot$s/kg and $\overline{\mathscr{Q}}_{M,1}=10^{5}$~kg/(m$^2$s) for the high one; the amplitudes of modulation are set to $\varepsilon_{Q_M,1}=\varepsilon_{Q_C,1}=0.9$ and the other parameters are unchanged.
The time evolution of the bulk energy is shown for the microstructured medium (plain line) and for the leading-order effective medium (dotted line) for these cases. Good agreement is obtained in each case, which validates the homogenisation of the dissipative terms. Additionally, one observes the competition between Floquet amplification and intrinsic losses. In the low dissipation case, the parametric amplification is still observed, but with a smaller rate than in Figure \ref{fig:leading_order_DD}(e). Using high dissipation, a decrease of energy is observed. It indicates that an optimal value of dissipation interface parameters may exist, where losses compensate exactly the parametric amplification. Such an equilibrium is investigated in \cite{TorrentPRB2018}.
}
\vspace{-10pt}
\paragraph{Impedance matching.}When the interface properties of the microstructured medium are chosen to ensure impedance matching, no reflections occur in the microstructured field. In \cite{Darche2025}, it is proven that the impedance matching condition for one interface {sinusoidally modulated} is  
\begin{equation}
    \label{eq:impedance_matching}
    \overline{\mathscr{M}}_1 = (\rho c)^2 \overline{\mathscr{C}}_1 \quad \text{ and } \quad \varepsilon_{M,1}=\varepsilon_{C,1}.
\end{equation}
By setting $\overline{\mathscr{K}}_1=1$~GPa/m, $\overline{\mathscr{M}}_1=11289.6$~kg/m$^{2}$,  $\varepsilon_{C,1}=\varepsilon_{M,1}=0.9$, this local impedance matching condition \eqref{eq:impedance_matching} is satisfied. Comparisons of the displacement fields for the effective medium and the microstructured one are presented in Figure \ref{fig:impedance} for $f_c=10$~Hz and $f_m=20$~Hz. The macroscopic behaviour is well accounted for by the effective field and the dispersive effects due to time modulations do not show up in that case. 
\vspace{-15pt}
\paragraph{The case of several modulated interfaces.}As the effective models have been developed for an arbitrary number $N$ of interfaces, we also present results for two interfaces within a unit cell. More precisely, they are located at $X_{n,1}=n\,h$ and $X_{n,2}=n\,h+0.65\,h$ and  described by, 
$$
\left\{
\begin{array}{lllll}
\overline{\mathscr{M}}_1=10^4~\mbox{kg/m}^{2}, & \overline{\mathscr{K}}_1=2.45~\mbox{GPa/m}, &\varepsilon_{C,1}=-0.9, &\varepsilon_{M,1}=0.9, &\Phi_{C,1}=\Phi_{M,1}=0,\\ [6pt]
\overline{\mathscr{M}}_2=2\times 10^4~\mbox{kg/m}^{2}, & \overline{\mathscr{K}}_2=1~\mbox{GPa/m}, &\varepsilon_{C,2}=0.5, &\varepsilon_{M,2}=0.5, &\Phi_{C,2}=\Phi_{M,2}=-\pi/2.
\end{array}
\right.
$$
The corresponding fields are presented in Figure \ref{fig:2_interf} for $f_c=10$~Hz and $f_m=20$~Hz (i.e $\eta_0=0.640$) and $f_m=20$~Hz (i.e.\ $\eta_1=1.92$). One observes again the same very good agreement between the exact field and its effective counterpart.

\subsection{High-order effective model }

\begin{figure}
    \centering
    \includegraphics[width=0.38\linewidth]{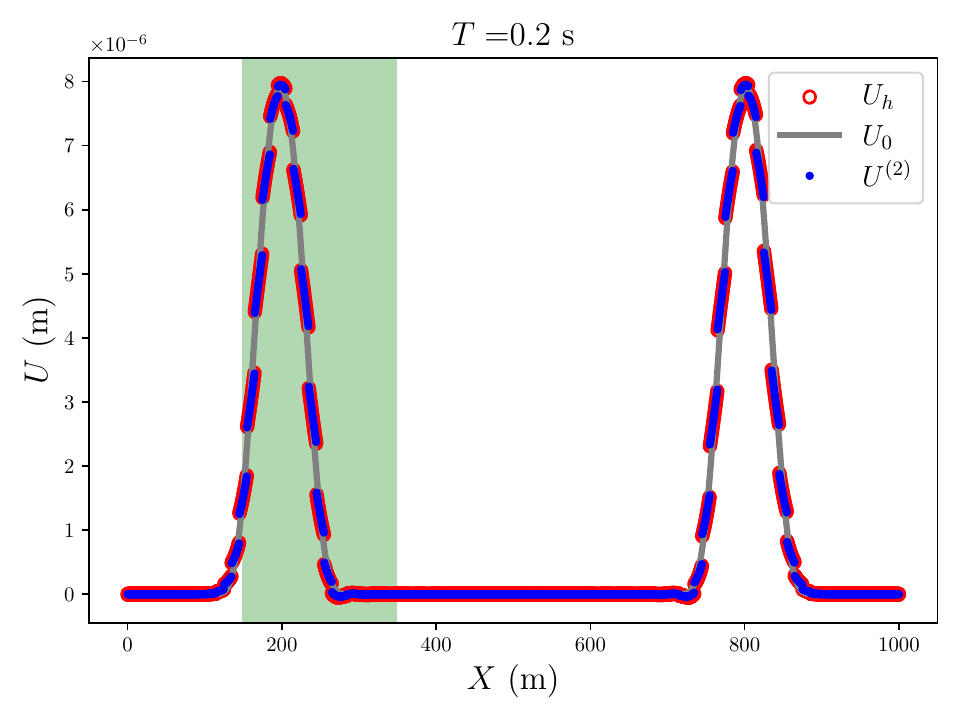}
    \includegraphics[width=0.38\linewidth]{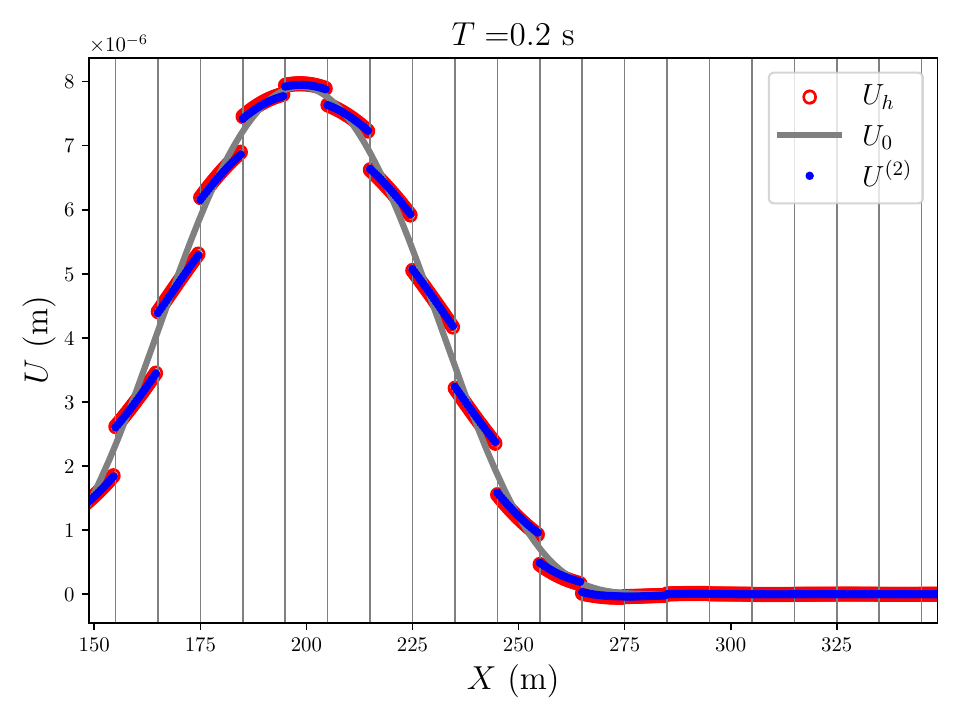} \\
    \includegraphics[width=0.38\linewidth]{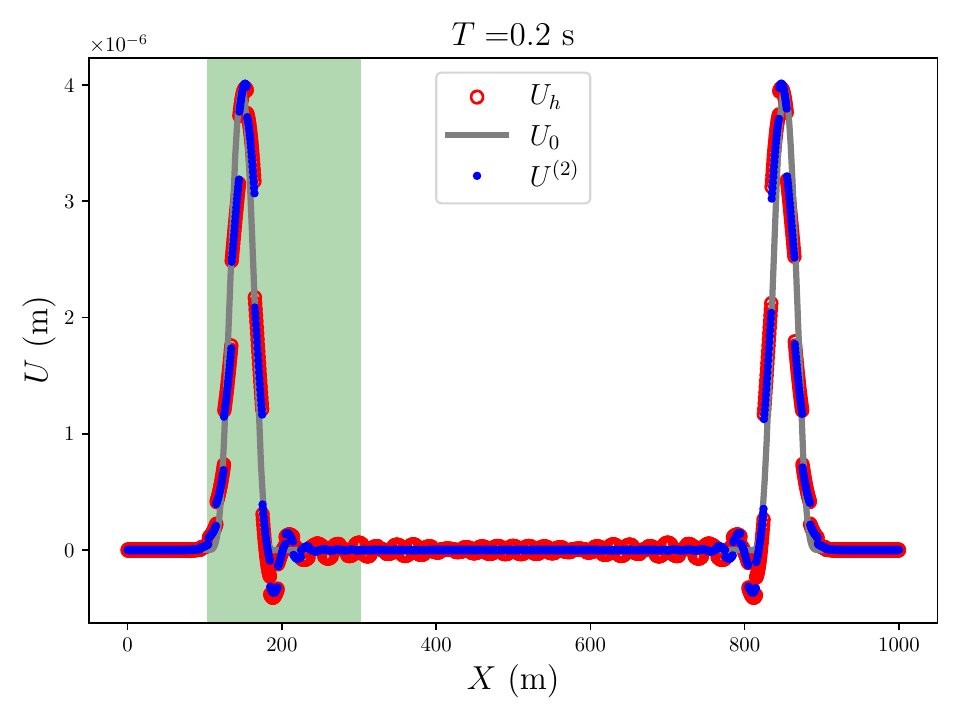}  
    \includegraphics[width=0.38\linewidth]{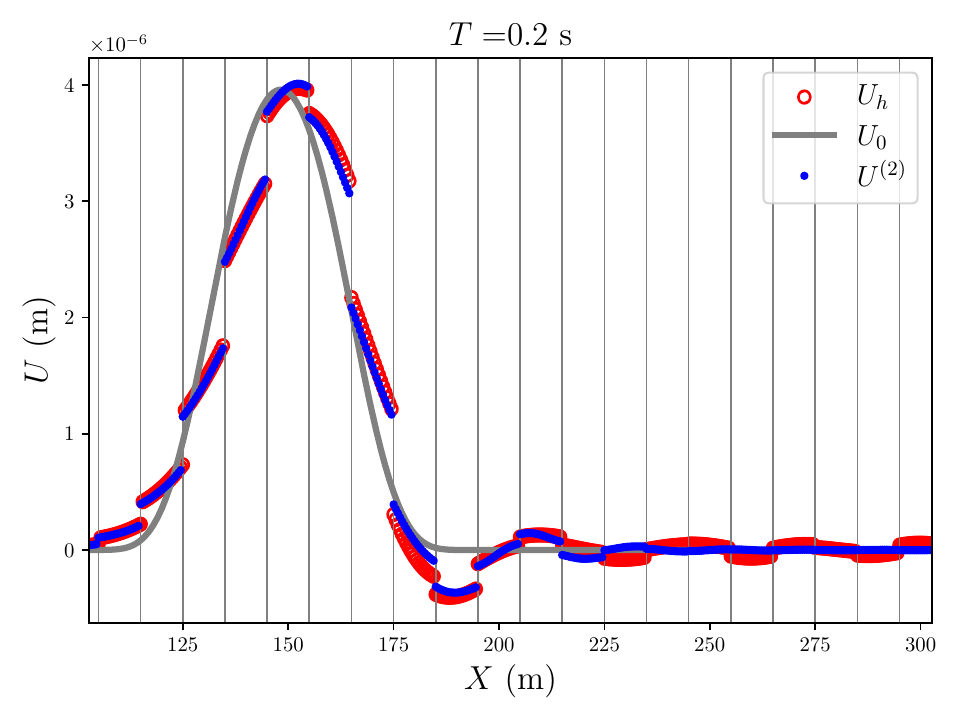} \\
    \includegraphics[width=0.38\linewidth]{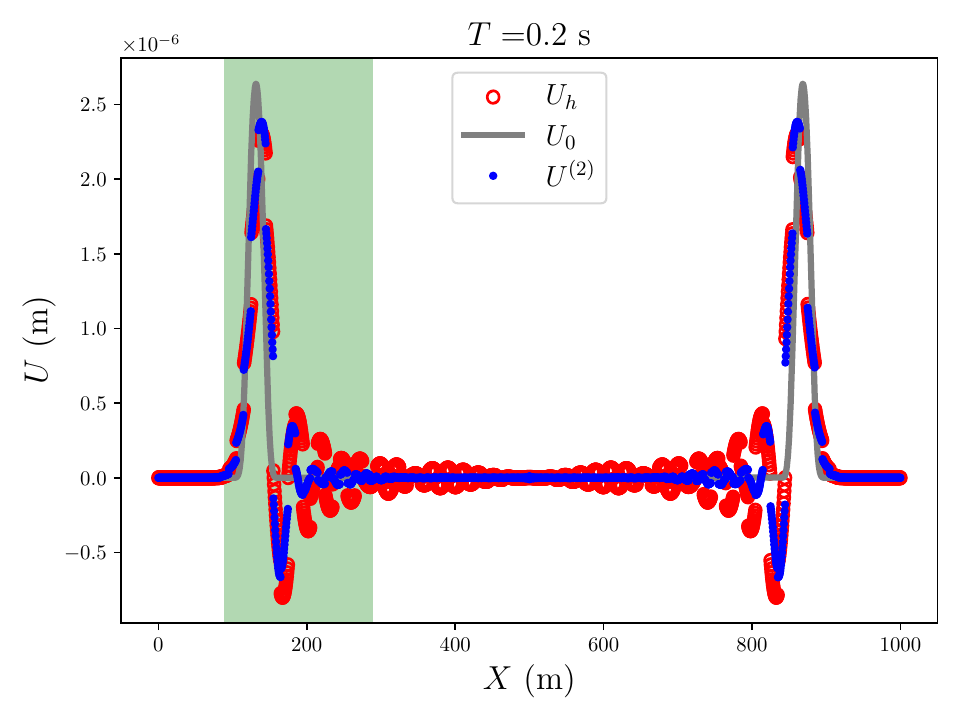} 
    \includegraphics[width=0.38\linewidth]{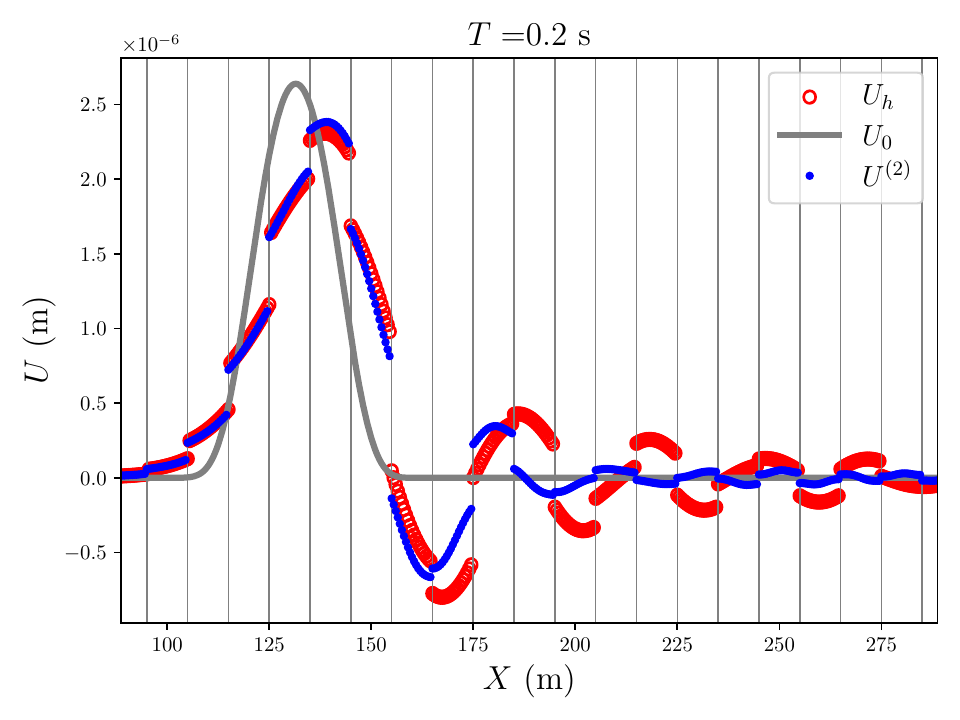} 
    \caption{Comparaison betweeen $U_h$ (full-field simulation in the microstructure), and the homogenised fields $U_0$ and $U^{(2)}$ without modulation ($f_m=0$) for an increasing value of $\eta_0$. From top to bottom : $f_c=\{10,20,30\}$~Hz corresponding to $\eta_0=\{0.31,0.63,0.94\}$.}
    \label{fig:increasing_fc}
\end{figure}

We now focus on the second-order effective model, and its accuracy with respect to the leading-order one. We consider a configuration with no mass for the interfaces, i.e.\ $\mathscr{M}_\ell=0$, for which the second-order model satisfies the simplified effective equation \eqref{eq:dim_order2}. More precisely, we take one single interface per interface with interface parameters $\overline{\mathscr{K}}_1=1$~{GPa/m} and $\varepsilon_{C,1}=0.9$. 
\vspace{-10pt}
\paragraph{High-order dispersive effects with unmodulated interfaces: $f_m=0$.}We start with an example without time-modulation of the interfaces which corresponds to \cite{bellisJotMaPoS2021}, where homogenisation was performed in a non-linear case but limited to the first order. We increase the central frequency $f_c$ of the source, and consequently the associated value of $\eta_0$. Comparisons of the displacement in the microstructured medium, with the leading-order and second-order effective displacements, are presented in Figure \ref{fig:increasing_fc}. For the lowest value of $\eta_0$, we do not see any difference on the mean value (macroscopic behaviour) between the exact, leading-order or second-order field. As one increases the central frequency, and as $\eta_0$ approaches 1, we get a shift of the dominant peak, which is missed by the leading-order model but well described by the second-order one. Moreover, the {corrector of the} second-order field also describes local fluctuations which are not described by the leading-order model. However, despite better describing dispersive effects, even the second-order model does not fully capture the larger oscillations that appear for the highest value of $\eta_0$ considered.

\begin{figure}
    \centering
    \includegraphics[width=0.38\linewidth]{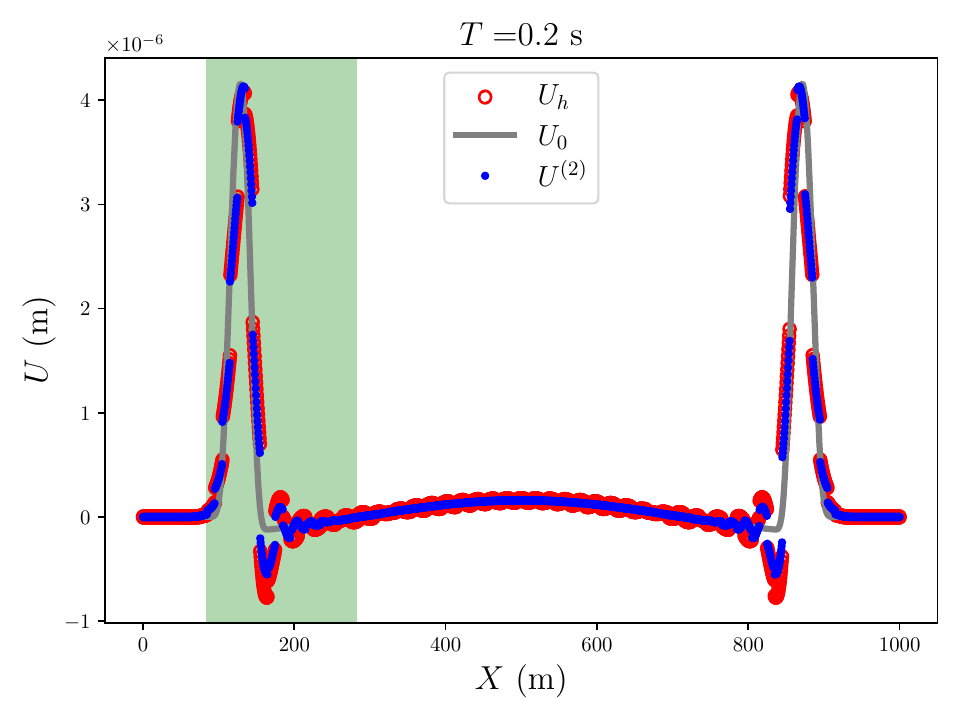} 
    \includegraphics[width=0.38\linewidth]{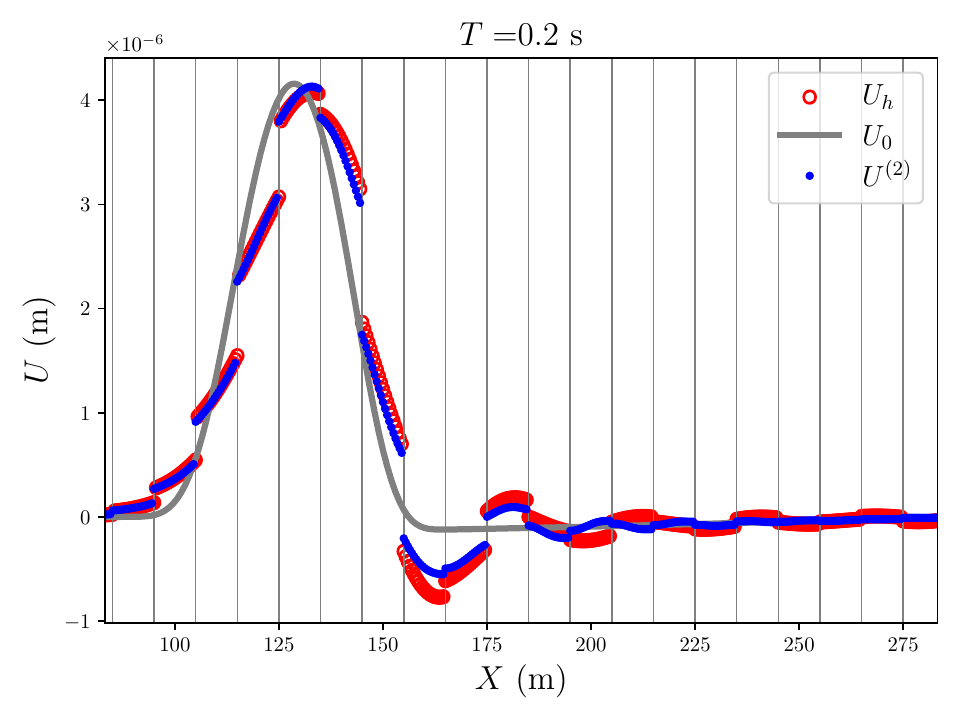} \\
    \includegraphics[width=0.38\linewidth]{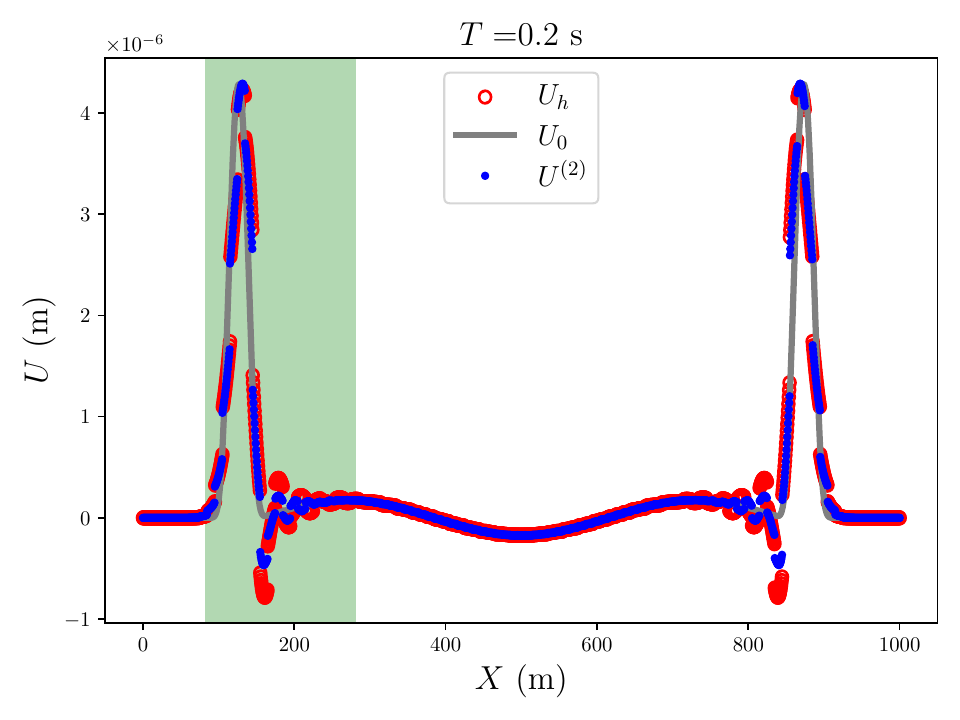} 
    \includegraphics[width=0.38\linewidth]{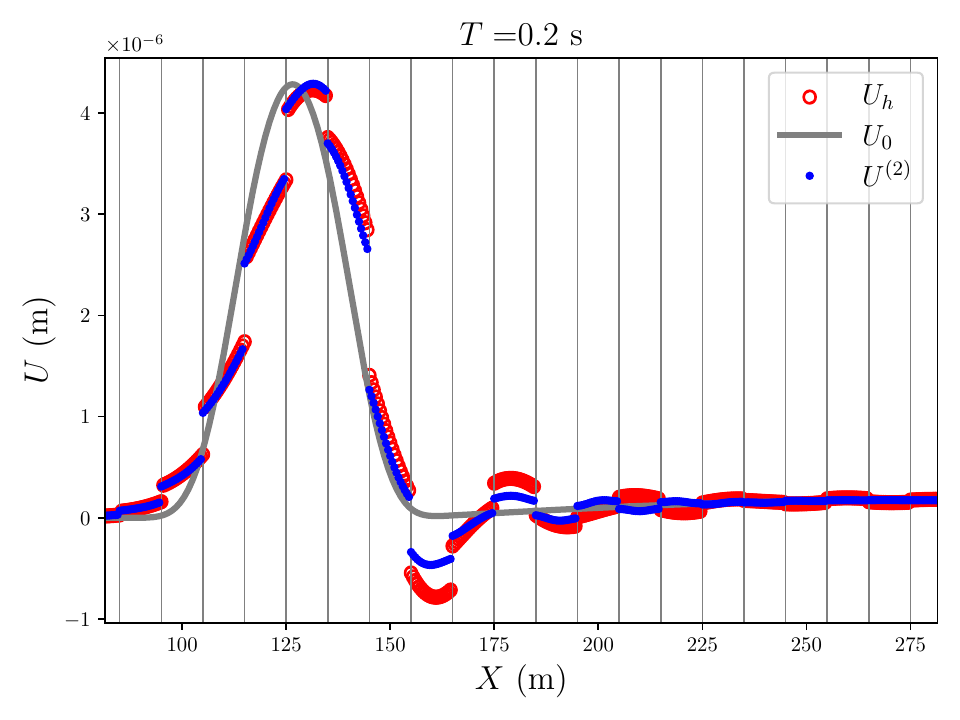} \\
    \includegraphics[width=0.38\linewidth]{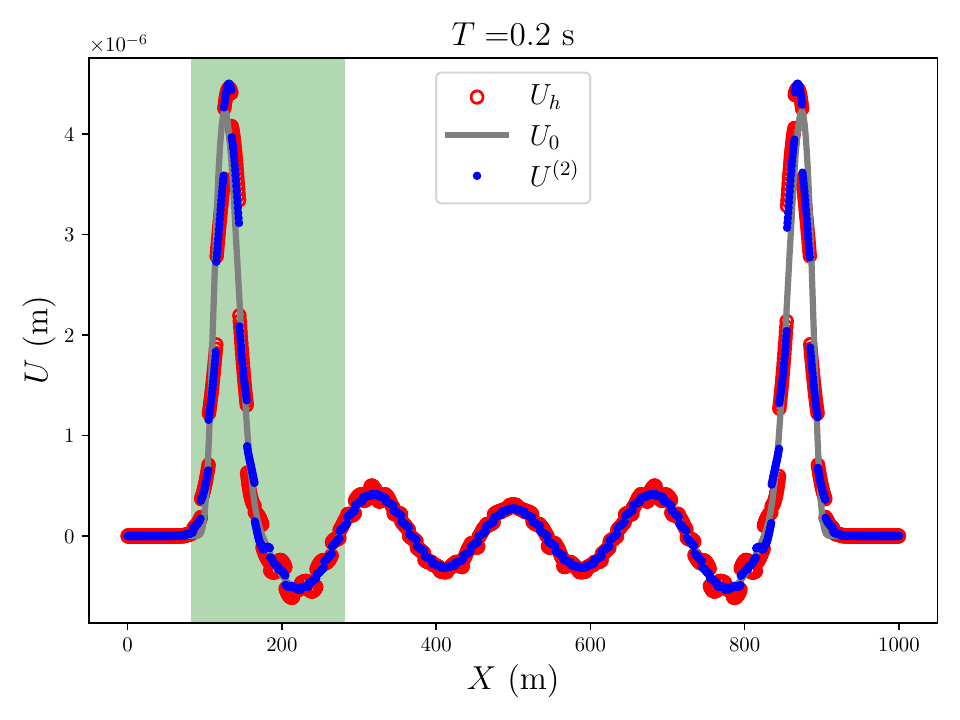} 
    \includegraphics[width=0.38\linewidth]{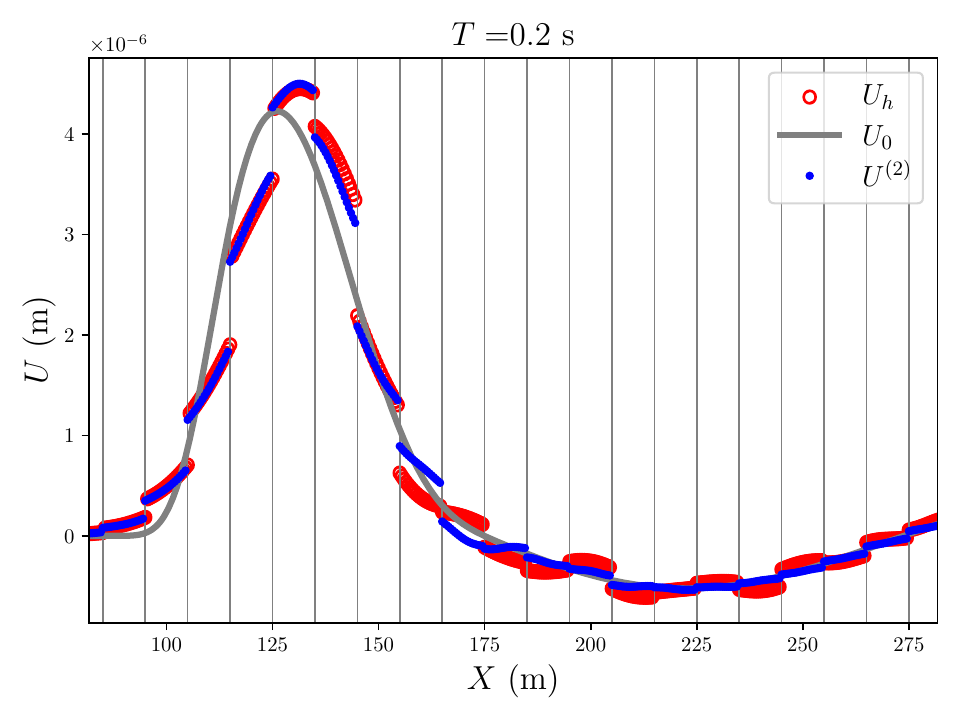}\\
    \includegraphics[width=0.38\linewidth]{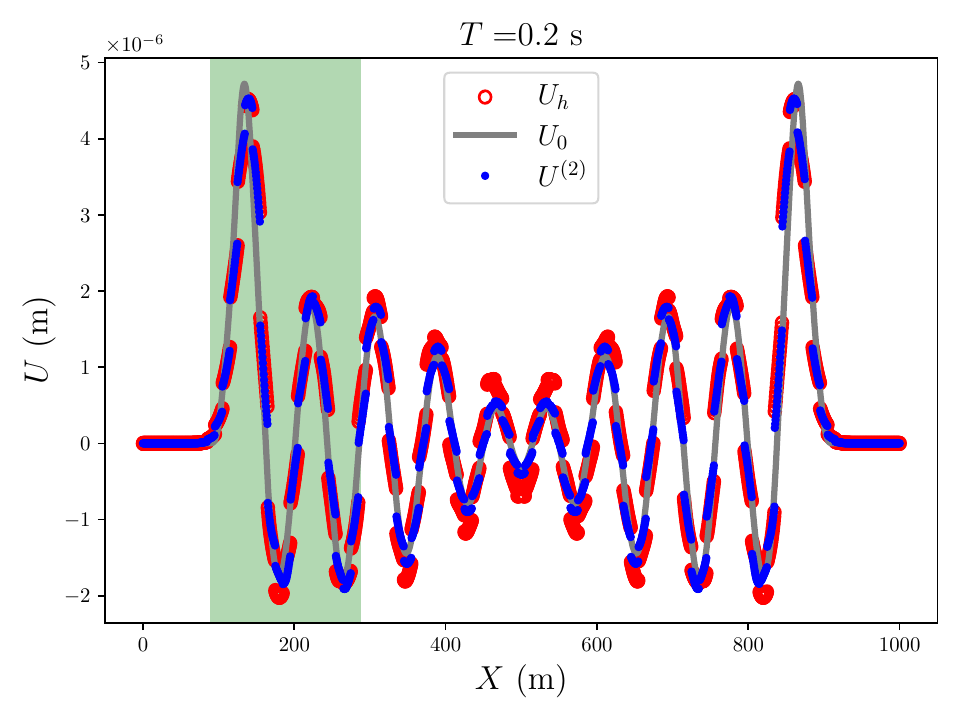} 
    \includegraphics[width=0.38\linewidth]{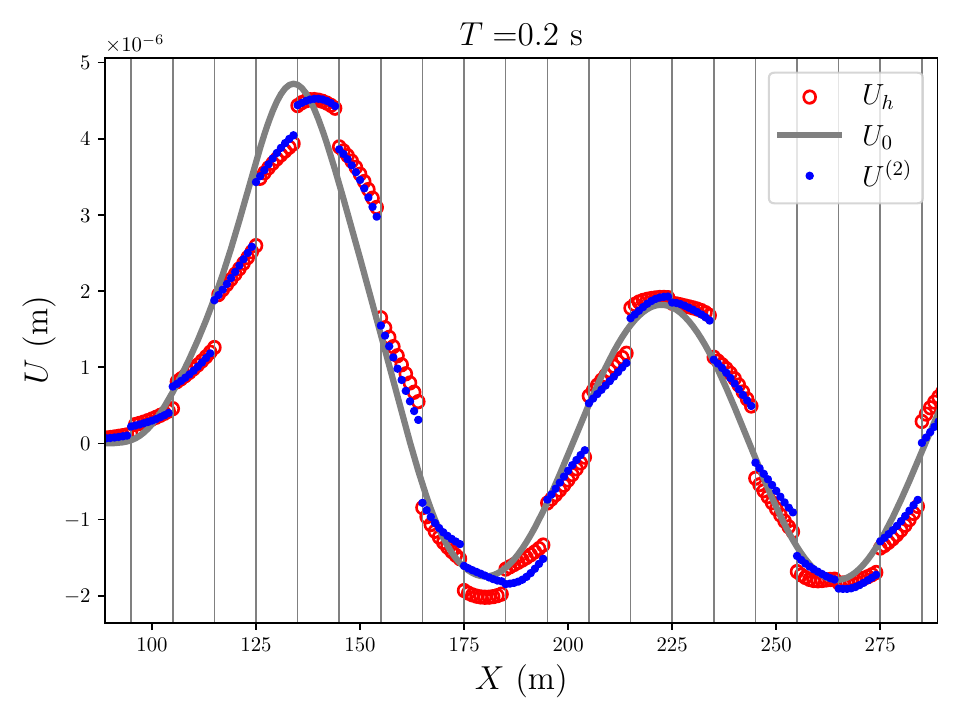} 
    \caption{Comparaison betweeen $U_h$ (full-field simulation in the microstructure), and the homogenised fields $U_0$ and $U^{(2)}$ with source pulse at $f_c=20$~Hz ($\eta_0=0.63)$. From top to bottom $f_m=\{5,10,20,50\}$~Hz i.e.\ $\eta_1=\{0.78,0.94,1.25,2.19\}$.}
    \label{fig:increasing_fm}
\end{figure}

\vspace{-10pt}
\paragraph{Intrinsic dispersion due to time-modulation: $f_c$ constant, increasing $f_m$.}We now set $f_c=20$~Hz, i.e.\ $\eta_0=0.63$ for which we had a good agreement without time-modulation of the interfaces, and we increase the frequency of modulation $f_m$ and consequently the value of the associated $\eta_1$. As shown in Figure \ref{fig:increasing_fm}, as we increase $f_m$, we get oscillations of larger amplitudes due to the time-modulation of the interfaces. These oscillations are very accurately accounted for by the effective models even at leading order. As previously, the second-order effective model accounts for the local fluctuations and the shift of the dominant peak while it is missed by the leading-order model. It is also interesting to notice the competition between high-order dispersive effects which exist even without time-modulation and the intrinsic dispersion due to the time-modulation. When the second one seems dominant, our effective models present a very good agreement even if very large values of $\eta_1$ are involved. 
\vspace{-10pt}

\begin{figure}
    \centering
    \includegraphics[width=0.4\linewidth]{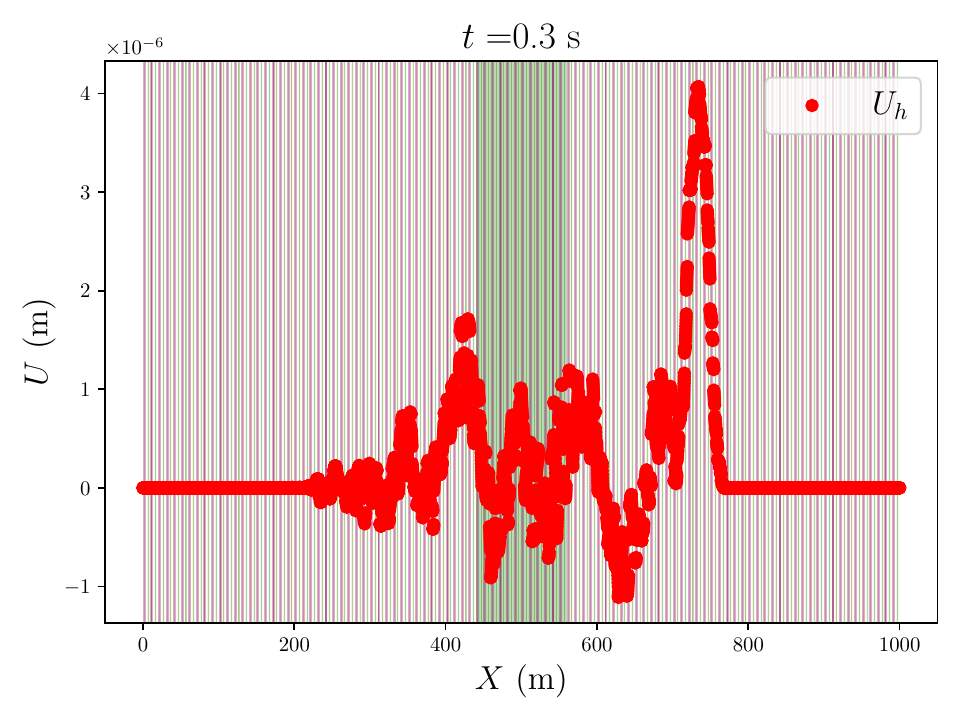}
    \includegraphics[width=0.4\linewidth]{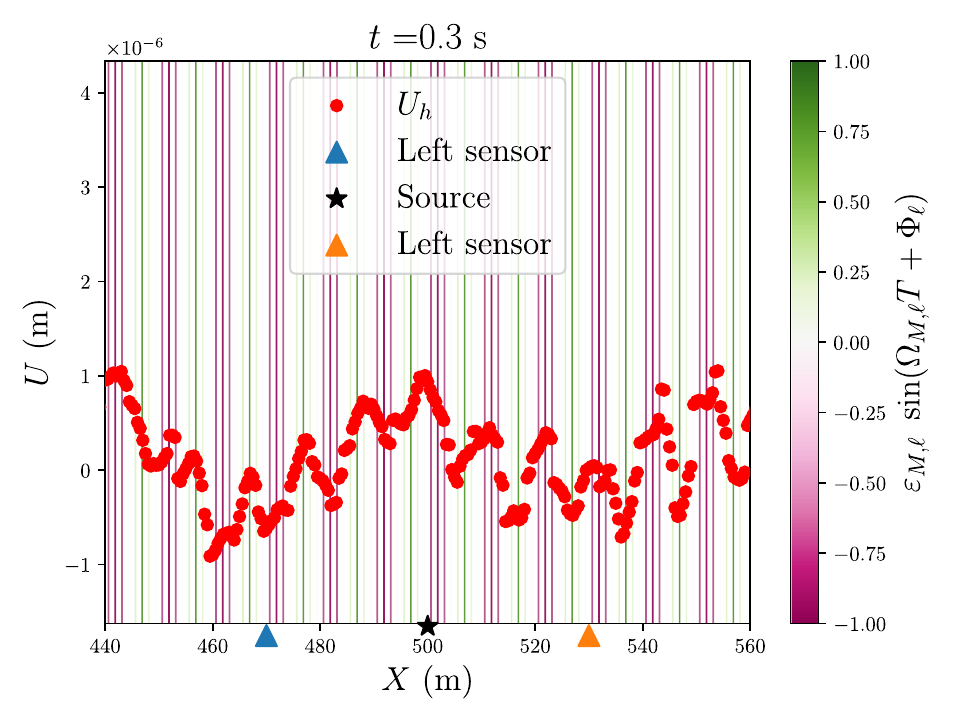}\\
    \includegraphics[width=0.4\linewidth]{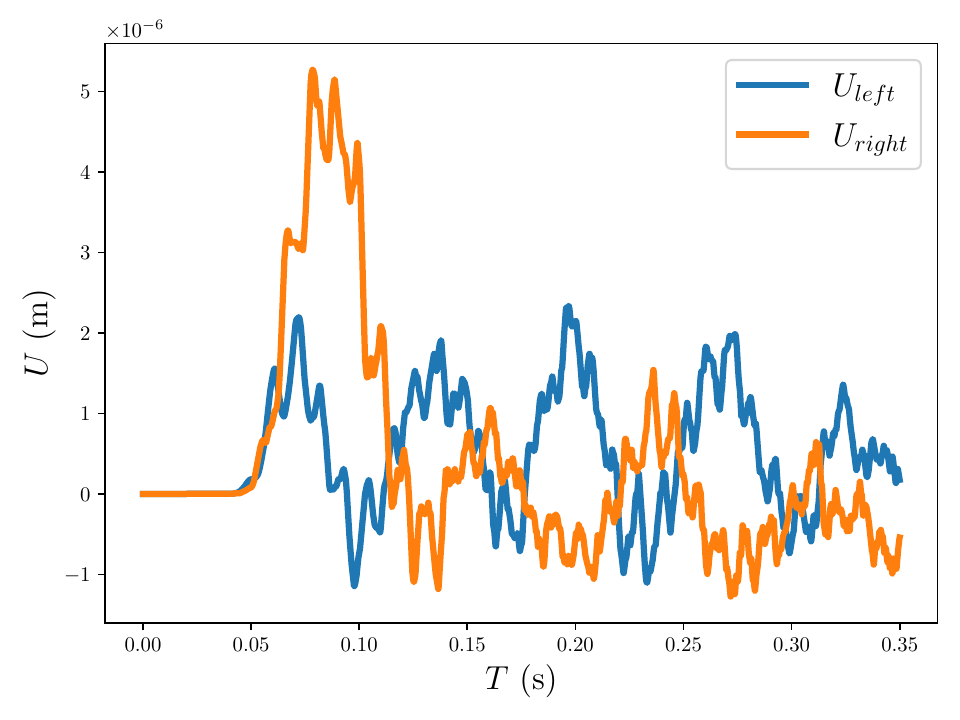}
    \caption{Observation of non-reciprocity for fast modulations: $N=8$ interfaces are modulated sinusoidally with a phase-shift (the coloured vertical lines denote the state of the interface at a given time). Top row: field $U_h$ in the whole microstructured medium where a clear asymmetry is observed (left) and zoom around a few unit cells (right).  Bottom row: Time history of $U_h$ measured at a receiver for two symmetric configurations where we swapped source {at $X=500$~m (black star)}  and receiver {at $X=470$~m (blue triangle) and $X=530$~m (orange triangle)}. Simulations done for $f_c=10$~Hz and $f_m=80$~Hz ($\eta_0=0.793$ and $\eta_1=7.14$).}
    \label{fig:NonReciprocity}
\end{figure}

\paragraph{Beyond the validity of our model: very fast modulations, and non-reciprocity. }
 We now consider $N=8$ interfaces {uniformly spaced} in a unit cell, which requires refining the meshing; in practice we use $N_X=8000$ grid nodes. Every interface is defined by $\mathscr{K}_\ell=4 \, \text{GPa/m}$, ${\mathscr{M}_\ell=5000 \, \text{kg/m}^{2}}$, $\varepsilon_{C,\ell}=-0.9$, $\varepsilon_{M,\ell}=0.9$ but with $\Phi_{C,\ell}=\Phi_{M,\ell}{=-\frac{2\pi \ell}{N}}$, which reproduces a wave-like modulation of the interfaces. The different colors of the vertical lines in Figure \ref{fig:NonReciprocity} denotes the different states in which the interfaces are. Furthermore, the central frequency of the source is $f_c=10$~Hz while the modulation frequency is $f_m=80$~Hz, leading to parameters $\eta_0=0.793$ and $\eta_1=7.14$. Because of the very fast modulation of the interface parameters, $\eta_1$ is very large compared to 1 and this setting is beyond the limit of validity of the effective models. By computing the displacement in the microstructured medium, one notices in Figure \ref{fig:NonReciprocity} a very clear asymmetry between the left-going and right-going waves, which is confirmed by the fact that the signals registered at a source and a receiver are very different when we swap this source and this receiver. This setting is therefore clearly non-reciprocal, which motivates the future development of new effective models, {tailored to high modulation frequencies,} to describe these phenomena. 


\vspace{-10pt}
\section{Conclusion}

We have studied the properties of a one-dimensional array of time-modulated interfaces using time-dependent spring-mass {jump} conditions. We have focused on the long-wavelength (low-frequency) homogenisation regime, wherein time-modulation is slow, making use of two-scale asymptotic expansions. 

At leading order, we have obtained a time-dependent wave equation 
with effective mass density and Young's modulus which are homogeneous in space but depend on time. The occurrence of $k$-gaps in case of a periodic modulation has been analysed. Second-order homogenisation has allowed us to describe higher-order dispersive effects and local fluctuations. These findings have been illustrated through time-domain simulations, showing excellent agreement between full-field solutions and solutions of the effective models for moderate modulation frequencies, even for quite high central frequencies.

Neither the leading-order \eqref{leading_order_dim} nor the second-order \eqref{WaveTotal2} effective wave equations exhibit odd-order spatial derivatives. Consequently, spatial inversion symmetry is preserved, and the effective {models} obtained are reciprocal.  
{However, as we illustrated through direct time-domain numerical simulations, when considering very fast time modulation of the interfaces, the microstructured media clearly exhibits non-reciprocity. This regime lies beyond the validity of our current approach, but it clearly motivates further developments.}
 \vskip6pt

\enlargethispage{20pt}

\section*{Acknowledgments}
M. D. was funded as a post-doctoral researcher by the Institut M\'ecanique et Ing\'enierie (Marseille, France). S. G. was funded by UK Research and Innvovation (UKRI) under the UK government's Horizon Europe funding guarantee (grant number 10033143).

\appendix

\section{Leading-order homogenisation in the case with dissipation}\label{Sec:Homog_Dissip}

In this appendix, we consider the more general case \eqref{JCmicro} with dissipation, i.e.\ $\mathscr{Q}_{C,\ell}\neq 0$ and $\mathscr{Q}_{M,\ell}\neq 0$, {which requires a velocity-stress formulation}. The strategy being very similar, we underline only the differences with the case treated in Section~\ref{SecLeading}. 


\subsection{Two-scale analysis}\label{SecHomog2scalevs}

\paragraph{\bf Non-dimensionalisation.}With dissipation, it is necessary to introduce the non-dimensionalised velocity $v_\eta=\partial_t u_\eta$, and the non-dimensionalised dissipation interface parameters
\begin{equation}
\mathscr{q}_{c,\ell}=\frac{1}{\eta}\sqrt{E^\star\rho^\star}\mathscr{Q}_{C,\ell},\quad \mathscr{q}_{m,\ell}=\frac{1}{\eta}\frac{1}{\sqrt{E^\star\rho^\star}}\mathscr{Q}_{M,\ell}.
\label{KMadimvs}
\end{equation}
 The scaling in $1/\eta$ for $\mathscr{q}_{c,\ell}$ and  $\mathscr{q}_{m,\ell}$ is chosen so that the dissipation terms have a contribution of same order of magnitude than the mass and rigidity terms in the following non-dimensionalised jump conditions:
\begin{subnumcases}{\label{CascadeJCvs}}
\ds \jump{v_\eta}_{\ell}=\eta\,\partial_t\left(\mathscr{c}_\ell(t)\,\dmean{\sigma_\eta}_{\ell}\right)+\eta\,\mathscr{q}_{c,\ell}\dmean{\sigma_\eta}_{\ell},\label{CascadeJCvvs}\\ [-6pt]
\ds \jump{\sigma_\eta}_{\ell}=\eta\,\partial_t\left(\mathscr{m}_\ell(t)\,\dmean{v_\eta}_{\ell}\right)+\eta\,\mathscr{q}_{m,\ell}\dmean{v_\eta}_{\ell}.\label{CascadeJCSvs}
\end{subnumcases}

\vspace{-15pt}
\paragraph{\bf Asymptotic expansion.}The velocity-stress formulation being necessary to perform homogenisation in the presence of dissipation, the fields $v_\eta$ and $\sigma_\eta$ are expanded using the usual ansatz:
\begin{equation}
v_\eta(x,t)=\sum_{j\geq 0} \eta^j v_j(x,y,t) \text{ and } \sigma_\eta(x,t)=\sum_{j\geq 0} \eta^j \sigma_j(x,y,t).
\label{ansatzvs}
\end{equation}
For all $j\geq1$, the fields $v_j$ and $\sigma_j$ are assumed to be continuous with respect to the first variable and $1$-periodic with respect to the second variable, i.e.~$v_j(x,y,t)=v_j(x,y+1,t)$ for all $y$, and similarly for $\sigma_j$. 
Injecting the ansatz into  \eqref{EDPadim} yields {the following}PDEs
\begin{subnumcases}{\label{CascadePDEvs}}
\alpha(y)\,\partial_{t} v_\eta(x,y,t)=\left(\partial_x+\frac{1}{\eta}\,\partial_y \right)\sigma_\eta(x,y,t)+f(x,t),\label{CascadePDEvvs}\\[-2pt]
\partial_{t} \sigma_\eta(x,y,t)=\beta(y)\left(\partial_x+\frac{1}{\eta}\,\partial_y\right)v_\eta(x,y,t),\label{CascadePDEsvs}
\end{subnumcases}
together with time-dependent jump conditions \eqref{CascadeJCvs}.
 

\subsection{Model derivation}\label{SecNHomogOrdre0vs}

The ${\cal O}(\eta^{-1})$ terms in \eqref{CascadePDEvs} yield
\begin{equation}
\partial_y v_0=0,\qquad \partial_y \sigma_0=0,
\end{equation}
and the ${\cal O}(\eta^{0})$ terms in the jump conditions \eqref{CascadeJCvvs} and \eqref{CascadeJCSvs} lead to 
\begin{equation}
\jump{v_0}_\ell=0,\qquad \jump{\sigma_0}_\ell=0.
\end{equation}
Consequently one gets 
\begin{equation}
\label{macro_v0_sigma_0vs}
v_0(x,y,t) = \overline{v}_0(x,t) \text{ and } \sigma_0(x,y,t)=\overline{\sigma}_0(x,t).
\end{equation}
The ${\cal O}(\eta^{0})$ term in \eqref{CascadePDEsvs} reads
\begin{equation}
\label{dt_sigma0vs}
\partial_t\sigma_0 = \beta\partial_xv_0+\beta\partial_yv_1.
\end{equation}
Differentiation of this equation with respect to $y$, together with \eqref{macro_v0_sigma_0vs}, yields
\begin{equation}
\partial_y(\beta\partial_yv_1) =-\beta'(y)\partial_xv_0.
\end{equation}
This equation on $v_1$ comes with the following jump conditions 
\begin{equation}{\label{JC_v1vs}}
\ds \jump{v_1}_{\ell}=(\mathscr{c}_\ell'+\mathscr{q}_{c,\ell})\sigma_0+\mathscr{c}_\ell\beta(\partial_xv_0+\partial_yv_1),\qquad 
\ds \jump{\beta(\partial_xv_0+\partial_yv_1)}_{\ell}=0.
\end{equation}
The first one is obtained by identifying the ${\cal O}(\eta)$ term in \eqref{CascadeJCvvs} and using \eqref{dt_sigma0vs}, and the second one by taking the jump of \eqref{dt_sigma0vs} while remembering that $\jump{\sigma_0}_\ell=0$.
Consequently $v_1$ can be written as 
\begin{equation}
\label{form_v1}
v_1(x,y,t)=\overline{v}_1(x,t)+P_1(y,t)\partial_xv_0(x,t)+Q(y,t)\sigma_0(x,t),
\end{equation}
where $P_1$ has already been introduced without dissipation \eqref{pb_cell_P1} and $Q$ is solution of the following cell problem:
\begin{equation}
(\mathscr{Q}):
\left|\begin{array}{ll}
\ds \partial_y(\beta \partial_yQ)=0 , & \text{ on } \bigcup\limits_{\ell=1}^{N}(y_{\ell},y_{\ell+1}) \\ [-2pt]
\ds \jump{\beta\partial_yQ}_\ell = 0, & \text{ for } \ell=1,\cdots,N\\ [-2pt]
 \ds\jump{Q}_\ell = \mathscr{c}_\ell'(t)+\mathscr{q}_{c,\ell}+\mathscr{c}_\ell(t)\,\beta\partial_yQ & \text{ for } \ell=1,\cdots,N\\[-2pt]
\ds \mean{Q}=0,\quad Q\,\mbox{ 1-periodic in }y.&
\end{array}
\right.
\label{pb_cell_P2vs}
\end{equation}
As $P_1$, $Q$ also depends on $t$ because of the time-modulation of the interface compliance. 

To obtain the leading-order model, one averages the ${\cal O}(\eta^{0})$ terms in \eqref{CascadePDEvvs} while using \eqref{meandy_jump} and the terms of order ${\cal O}(\eta^{1})$ in \eqref{CascadeJCSvs}, and averages \eqref{dt_sigma0vs} while using \eqref{form_v1}:
\begin{equation}
\label{effective_zero_non_dimvs}
\left\lbrace
\begin{aligned}
&\partial_t(\alpha_0(t)\overline{v}_0(x,t))+\gamma_{m,0}(t)\overline{v}_0(t) =\partial_x\overline{\sigma}_0(x,t)+f(x,t),\\[-2pt]
&\partial_t\overline{\sigma}_0(x,t)=\beta_0(t)\partial_x\overline{v}_0(x,t)+\gamma_{c,0}(t)\overline{\sigma}_0(t), 
\end{aligned}
\right.
\end{equation}
where $\alpha_0$ and $\beta_0$ are unchanged \eqref{def_effective_alpha_beta_gamma} but we have introduced the following effective parameters
\begin{equation}
\label{def_effective_alpha_beta_gammavs}
\gamma_{m,0}(t) = \sum_{\ell=1}^{N}\mathscr{q}_{m,\ell}(t), \qquad\gamma_{c,0}(t) = \mean{\beta\partial_yQ(\cdot,t)}.
\end{equation}
One can look for an explicit expression for $\gamma_{c,0}(t)$ as we did for $\beta_0$ \eqref{Beta0} and get:
\begin{equation}
\gamma_{c,0}(t)= \beta\partial_y Q=-\sum_{\ell=1}^{N}[\mathscr{c}'_\ell(t)+\mathscr{q}_{c,\ell}]\beta_0(t).
\label{Gamma0Defvs}
\end{equation}

\vspace{-10pt}
\paragraph{\bf Limit case without dissipation.}If there is no dissipation, i.e.\ $\mathscr{q}_{c,\ell}=0$ and $\mathscr{q}_{m,\ell}=0$ for $\ell=1,\cdots,N$, then \eqref{effective_zero_non_dimvs} reads
\begin{equation}
\partial_t(\alpha_0(t)\overline{v}_0(x,t)) =\partial_x\overline{\sigma}_0(x,t)+f(x,t),\qquad \partial_t\overline{\sigma}_0(x,t)=\beta_0(t)\partial_x\overline{v}_0(x,t)+\gamma_{c,0}(t)\overline{\sigma}_0(t),
\end{equation}
where $\gamma_{c,0}(t)$ satisfies $\gamma_{c,0}(t) = \frac{\beta_0'(t)}{\beta_0(t)}.$ The second equation integrated in time reads $\overline{\sigma}_0(x,t) =\beta_0(t)\partial_x\overline{u}_0(x,t)$. Finally, the effective leading-order model can be reduced to \eqref{WaveAdim0}.
\vspace{-10pt}
\paragraph{\bf Dimensionalised coordinates.}From  \eqref{KMadimvs}, \eqref{Adim} and \eqref{AlphaBeta}, one obtains the dimensionalised leading-order model
\begin{equation}
\label{leading_order_dimvs}
\left\lbrace
\begin{aligned}
&\rho_0(T)\,\partial_T\overline{V}_0(X,T)=\partial_X\overline{S}_0(X,T)+\Gamma_{M,0}(T)\,\overline{V}_0(X,T)+F(X,T),\\
&\partial_T\overline{S}_0(X,T)=E_0(T)\,[\partial_X\overline{V}_0(X,T)+\Gamma_{C,0}(T)\,\overline{S}_0(X,T) ],
\end{aligned}
\right.
\end{equation}
together with the dimensionalised effective dissipation parameters: 
\begin{equation}
\Gamma_{M,0}(T)=-\frac{1}{h}\sum_{\ell=1}^{N}\left[\mathscr{M}_\ell'(T)+\mathscr{Q}_{M,\ell}(T) \right],\qquad \Gamma_{C,0}(T)=-\frac{1}{h}\sum_{\ell=1}^{N}\left[\mathscr{C}_\ell'(T)+\mathscr{Q}_{C,\ell}(T) \right].
\label{E0R0vs}
\end{equation}


\subsection{Numerical modelling}\label{SecNumDissip}

The velocity-stress formulation \eqref{E0R0vs} can be recast as: 
\begin{equation}
\partial_T\textbf{U}+\textbf{A}(T)\,\partial_X\textbf{U}=\textbf{S}(T)\,\textbf{U}+\textbf{F}(X,T),
\label{eq:HomoVS}
\end{equation}
where the material properties depend on time but are homogeneous in space. The vectors and matrices are expressed as:
\begin{equation}
\begin{array}{ll}
{\bf U}(X,T)=\left( 
\begin{array}{c}
\overline{V}_0(X,T)\\
\overline{S}_0(X,T)
\end{array} 
\right), \hspace{1cm} &
{\bf F}(X,T)=\left( 
\begin{array}{c}
\ds \frac{F(X,T)}{\rho_0(T)}\\
0
\end{array} 
\right), \hspace{0.5cm} \\ [12pt]
{\bf A}(T)=\left(
\begin{array}{cc}
0 & \ds -\frac{1}{\rho_0(T)}\\
-E_0(T) & 0
\end{array}
\right), & 
{\bf S}(T)=\left(
\begin{array}{cc}
\ds \frac{1}{\rho_0(T)}\Gamma_{M,0}(T)& 0\\
0 & E_0(T)\Gamma_{C,0}(T)
\end{array}
\right).
\end{array}
\label{SysOrdre1}
\end{equation}
The system \eqref{eq:HomoVS} can be solved by a splitting method \cite{benjaziaWM2014}, where (\ref{eq:HomoVS}) is divided into a propagative part:
\begin{equation}
  \partial_T{\bf U}+{\bf A}(T)\,\partial_X{\bf U}={\bf 0},
  \label{eq:HomoPropPart}
\end{equation}
and a diffusive/relaxation part:
\begin{equation}
  \partial_T{\bf U}={\bf S}(T){\bf U}+{\bf F}(X,T).
  \label{eq:HomoPropPartbis}
\end{equation}
The solutions to \eqref{eq:HomoPropPart} and \eqref{eq:HomoPropPartbis} are given by the time-dependent discrete operators $\textbf{H}_p$ and $\textbf{H}_d$, respectively. Second-order Strang's splitting is then used, leading to the time-marching
\begin{equation}
\label{StrangsSplitting}\textbf{U}^{(1)}_m=\mathcal{\textbf{H}}_d\left(T_n,\frac{\Delta T}{2}\right)\textbf{U}^n_m,\quad
\textbf{U}^{(2)}_m=\mathcal{\textbf{H}}_p\left(\Delta T\right)\textbf{U}^{(1)}_m,\quad
\textbf{U}^{n+1}_m=\mathcal{\textbf{H}}_d\left(T_{n+1},\frac{\Delta T}{2}\right)\textbf{U}^{(2)}_m.
\end{equation}
The operator $\mathcal{\textbf{H}}_p$ relies on the same ADER-4 scheme as for the microstructured medium (Section~\ref{Sec:Methods}). For the operator $\mathcal{\textbf{H}}_d$, an analytical solution yields: 
\begin{equation}
  \mathcal{\textbf{H}}_d\left(T,\frac{\Delta T}{2}\right){\bf U}_m=\exp\left(\int_{T}^{T+\Delta T/2} {\bf S(\tau)}d\tau\right)\,{\bf U}_m.
\end{equation}
During one time step, we consider that the parameters do not vary, so that \begin{equation}
   \mathcal{\textbf{H}}_d\left(T_n,\frac{\Delta T}{2}\right){\bf U}_m=\exp\left({\bf S}(T_n)\frac{\Delta T}{2}\right)\,{\bf U}_m.
\end{equation}


\section{Dispersion relation for the time-periodic leading-order model}\label{AppPWE}

We detail here the Plane Wave Expansion (PWE) methods used to derive the dispersion relation associated with the leading-order effective equation \eqref{leading_order_dim} in case of a time-periodic modulation of period $\tau$. The unit cell in time is denoted by $\mathcal{T}=[0,\tau]$. 
\vspace{-10pt}
\paragraph{\bf Fourier series expansion.}As the material properties and the function $U^\sharp$  are $\tau$-periodic, they can be expanded in Fourier series:
\begin{equation}
\rho_0(T)=\sum_{n\in\mathbb{Z}}\rho_ne^{\text{i}\frac{2\pi}{\tau}nT},\qquad E_0(T)=\sum_{n\in\mathbb{Z}}  E_n e^{\text{i}\frac{2\pi}{\tau}nT},\qquad
U^\sharp(T)=\sum_{n\in\mathbb{Z}}U^\sharp_ne^{\text{i}\frac{2\pi}{\tau}nT},
\label{SFpropT}
\end{equation}
where the coefficient $\rho_n$ and $E_n$ are :
\begin{equation}
  \rho_n=\frac{1}{\tau}\int_{\mathcal{T}}\rho_0(T)e^{-i\frac{2\pi}{\tau}nT}dT,\qquad
  E_n=\frac{1}{\tau}\int_{\mathcal{T}}E_0(T)e^{-i\frac{2\pi}{\tau}nT}dT.
\end{equation} 
Using the expression of $U^\sharp$ into \eqref{eq:UBlocht}, the leading-order homogenised displacement field can be written:
\begin{equation}
  \hat{U}(T)=\sum_{n\in\mathbb{Z}}U^\sharp_ne^{i(\frac{2\pi}{\tau}n+\omega^\sharp)\,T}.
\end{equation}
Injecting the Fourier series expansion of the displacement field and of the material parameters into the wave equation \eqref{eq:Wave1DFt}, one obtains
\begin{equation}
  \sum_{n\in\mathbb{Z}}\sum_{m\in\mathbb{Z}}\left(\rho_m\left(\frac{2\pi}{\tau}n+\omega^\sharp\right)\left(\frac{2\pi}{\tau}(n+m)+\omega^\sharp\right)U^\sharp_n-k^2E_mU^\sharp_n\right)e^{i\frac{2\pi}{\tau}(n+m)\,T}=0.
  \label{eq:FourierSET}
\end{equation}

\vspace{-10pt}
\paragraph{\bf Scalar product.}The hermitian scalar product is defined for $f$ and $g$ two $\tau$-periodic functions:
\begin{equation}
\left\langle f,g\right\rangle =\frac{1}{\tau}\int_{-\frac{\tau}{2}}^{\frac{\tau}{2}}f(T)\,\overline{g}(T)\,dT.
\end{equation}
where $\overline{g}(T)$ denotes the complex conjugate of $g$. Taking $f(T)=e^{i\frac{2\pi}{\tau}(n+m)T}$ and $g(T)=e^{i\frac{2\pi}{\tau}pT}$ yields:
\begin{equation}
  \left\langle e^{i\frac{2\pi}{\tau}(n+m)T},e^{i\frac{2\pi}{\tau}pT}\right\rangle=\frac{1}{\tau}\int_{-\frac{\tau}{2}}^\frac{\tau}{2}e^{-i\frac{2\pi}{\tau}(n+m-p)\,T}dT=\left\{\begin{array}{c}
    0\text{ if }n+m\neq p, \\
    1\text{ if }n+m=p.
  \end{array}
  \right.
\end{equation}
By taking the scalar product of the \eqref{eq:FourierSET} and the $\tau$-periodic function $g(T)$ previously defined, the following relation is obtained for each value of $p\in \mathbb{Z}$:
\begin{equation}
  \sum_{n\in\mathbb{Z}}\left(\frac{2\pi}{\tau}n+\omega^\sharp\right)\left(\frac{2\pi}{\tau}p+\omega^\sharp\right)\rho_{p-n}U^\sharp_n-k^2E_{p-n}U^\sharp_n=0,
  \label{eq:PSeq}
\end{equation}
where relation $m=p-n$ is used.  This equation is truncated to $n=-N_f,\cdots, N_f$, where $N_f$ is the number of Fourier modes, and we choose $p=-N_f,\cdots,N_f$.  It follows that the vector ${\bf U}=\left(U^\sharp_{-N_f}, ..., U^\sharp_{N_f}\right)^T$ satisfies the generalised eigenvalue problem 
\begin{equation}
  {\bm P(\omega^\sharp)\,{\bf U}}=k^2\,{\bm Q\,{\bf U}},
  \label{EigenGene}
\end{equation}
with the components:
\begin{equation}
P_{(N+n)(N+p)}=\left(\frac{2\pi}{\tau}n+\omega^\sharp\right)\left(\frac{2\pi}{\tau}p+\omega^\sharp\right)\rho_{p-n},\qquad 
Q_{(N+n)(N+p)}=E_{p-n}.
\label{eq:CoefsGEV}
\end{equation}

\begin{Remark}
The matrices $\bm{P}$ and $\bm{Q}$ in the PWE are hermitian. Consequently, the Generalised Eigenvalue Problem (\ref{EigenGene}) has real eigenvalues and its eigenvectors form a basis. One can easily show it by using the fact that the effective parameters are real. 
\end{Remark}


\section{Cell problems for the correctors in $u_2$}\label{AppCellU2}

Injecting the displacements $u_0$ \eqref{macro_v0_sigma_0}, $u_1$ \eqref{form_u1}, and $u_2$ \eqref{U2} into \eqref{JC_o2} and using the leading-order wave equation \eqref{WaveAdim0}, one obtains the cell problems for the three correctors in \eqref{U2}:

\begin{align}
&(\mathscr{P}_2):
\left|
\begin{array}{ll}
\ds \partial_y \left(\beta\left(P_1+\partial_y P_2\right)\right)=0, &\text{ on } \bigcup\limits_{\ell=1}^{N}(y_{\ell},y_{\ell+1})\\ [-2pt]
\ds \jump{P_2}_\ell=\mathscr{c}_\ell(t)\,\dmean{\beta\left(P_1+\partial_y P_2\right)}_\ell,  & \text{ for } \ell=1,\cdots,N\\ [-2pt]
\ds \jump{\beta\left(P_1+\partial_y P_2\right)}_\ell=0, & \text{ for } \ell=1,\cdots,N\\ [-2pt]
\ds \mean{P_2}=0,\quad P_2\,\mbox{ 1-periodic in }y,
\end{array}
\right.
\label{CellPbP2} \\
&(\mathscr{P}_3):
\left|
\begin{array}{ll}
\ds \partial_y \left(\beta\,\partial_y P_3\right)+\alpha_0(t)-\alpha=0, &\text{ on } \bigcup\limits_{\ell=1}^{N}(y_{\ell},y_{\ell+1})\\ [-2pt]
\ds \jump{P_3}_\ell=\mathscr{c}_\ell(t)\,\dmean{\beta\,\partial_y P_3}_\ell, & \text{ for } \ell=1,\cdots,N\\ [-2pt]
\ds \jump{\beta\,\partial_y P_3}_\ell=\mathscr{m}_\ell(t), & \text{ for } \ell=1,\cdots,N\\ [-2pt]
\ds \mean{P_3}=0,\quad P_3\,\mbox{ 1-periodic in }y,&
\end{array}
\right.
\label{CellPbP3} \\
&(\mathscr{P}_4):
\left|
\begin{array}{ll}
\ds \partial_y \left(\beta\,\partial_y P_4\right)+\alpha'_0(t)=0, &\text{ on } \bigcup\limits_{\ell=1}^{N}(y_{\ell},y_{\ell+1})\\ [2pt]
\ds \jump{P_4}_\ell=\mathscr{c}_\ell(t)\,\dmean{\beta\,\partial_y P_4}_\ell, & \text{ for } \ell=1,\cdots,N\\ [-2pt]
\ds \jump{\beta\,\partial_y P_4}_\ell=\mathscr{m}'_\ell(t),  & \text{ for } \ell=1,\cdots,N\\ [-2pt]
\ds \mean{P_4}=0,\quad P_4\,\mbox{ 1-periodic in }y,&
\end{array}
\right.
\label{CellPbP4}
\end{align}
where $\alpha_0$ is defined in \eqref{def_effective_alpha_beta_gamma}. Integration of \eqref{CellPbP2} on $(y_{\ell},y_{\ell+1})$  gives
\begin{equation}
\beta\left(P_1+\partial_y P_2\right)=\mathscr{C}_{\partial P_2,\ell}(t),\qquad \ell=1,\cdots, N.
\end{equation}
The continuity of flux in \eqref{CellPbP2} ensures that $\mathscr{C}_{\partial P_2,\ell}:=\mathscr{C}_{\partial P_2}$ does not depend on $\ell$, and hence
\begin{equation}
\partial_y P_2=\frac{1}{\beta}\mathscr{C}_{\partial P_2}-P_1.
\end{equation}
Averaging gives
\begin{equation}
-\sum_{\ell =1}^{N} \jump{P_2}_\ell=\mean{\frac{1}{\beta}}\mathscr{C}_{\partial P_2}-\mean{P_1},
\end{equation}
and hence $\mathscr{C}_{\partial P_2}=0.$ It follows
\begin{equation}
P_1+\partial_y P_2=0.
\label{P1dP2}
\end{equation}

\section{Cell problems for the correctors in $u_3$}\label{AppCellU3}

Injecting \eqref{U3} into \eqref{Ordre2-A} and into the jump conditions \eqref{EDP-eta1b} and \eqref{EDP-eta1c}, and using \eqref{P1dP2}, one obtains the cell problems for the new four correctors in \eqref{U3}: 
\begin{align}
&(\mathscr{P}_5):
\left|
\begin{array}{ll}
\ds \partial_y(\beta\,(P_2+\partial_y P_5))=0, &\text{ on } \bigcup\limits_{\ell=1}^{N}(y_{\ell},y_{\ell+1})\\ [-2pt]
\ds \jump{P_5}_\ell=\mathscr{c}_\ell(t)\,\dmean{\beta\left(P_2+\partial_y P_5\right)}_\ell,  & \text{ for } \ell=1,\cdots,N\\ [-2pt]
\ds \jump{\beta\left(P_2+\partial_y P_5\right)}_\ell=0, & \text{ for } \ell=1,\cdots,N\\ [-2pt]
\ds \mean{P_5}=0,\quad P_5\,\mbox{ 1-periodic in }y,&
\end{array}
\right.
\label{CellPbP5} \\
&(\mathscr{P}_6):
\left|
\begin{array}{ll}
\ds \partial_y(\beta\left(P_3+\partial_y P_6\right))+\beta\,\partial_y P_3-\alpha\,P_1=0, &\text{ on } \bigcup\limits_{\ell=1}^{N}(y_{\ell},y_{\ell+1})\\ [-2pt]
\ds \jump{P_6}_\ell=\mathscr{c}_\ell(t)\,\dmean{\beta\left(P_3+\partial_y P_6\right)}_\ell, & \text{ for } \ell=1,\cdots,N\\ [-2pt]
\ds \jump{\beta\left(P_3+\partial_y P_6\right)}_\ell=\mathscr{m}_\ell(t)\,\dmean{P_1}_\ell,& \text{ for } \ell=1,\cdots,N\\ [-2pt]
\ds \mean{P_6}=0,\quad P_6\,\mbox{ 1-periodic in }y,&
\end{array}
\right.
\label{CellPbP6}
\end{align}
\begin{align}
&(\mathscr{P}_7):
\left|
\begin{array}{ll}
\ds \partial_y(\beta\,(P_4+\partial_y P_7))+\beta\partial_y P_4-2\,\alpha\,\partial_t P_1=0, &\text{ on } \bigcup\limits_{\ell=1}^{N}(y_{\ell},y_{\ell+1})\\ [-2pt]
\ds \jump{P_7}_\ell=\mathscr{c}_\ell(t)\,\dmean{\beta\left(P_4+\partial_y P_7\right)}_\ell, & \text{ for } \ds \ell=1,\cdots,N\\ [-2pt]
\ds \jump{\beta\left(P_4+\partial_y P_7\right)}_\ell=\mathscr{m}'_\ell(t)\,\dmean{P_1}_\ell+2\,\mathscr{m}_\ell(t)\,\dmean{\partial_t P_1}_\ell,\qquad & \text{ for } \ell=1,\cdots,N\\ [-2pt]
\ds \mean{P_7}=0,\quad P_7\,\mbox{ 1-periodic in }y,&
\end{array}
\right.
\label{CellPbP7} \\
&(\mathscr{P}_8):
\left|
\begin{array}{ll}
\ds \partial_y(\beta\,\partial_y P_8)-\alpha\partial^2_{tt}P_1=0, &\text{ on } \bigcup\limits_{\ell=1}^{N}(y_{\ell},y_{\ell+1})\\ [-2pt]
\ds \jump{P_8}_\ell=\mathscr{c}_\ell(t)\,\dmean{\beta\,\partial_y P_8}_\ell, & \ds \text{ for } \ell=1,\cdots,N\\ [-2pt]
\ds \jump{\beta\,\partial_y P_8}_\ell=\mathscr{m}'_\ell(t)\,\dmean{\partial_t P_1}_\ell+\mathscr{m}_\ell(t)\,\dmean{\partial^2_{tt} P_1}_\ell,\qquad & \text{ for } \ell=1,\cdots,N\\ [-2pt]
\ds \mean{P_8}=0,\quad P_8\,\mbox{ 1-periodic in }y.&
\end{array}
\right.
\label{CellPbP8}
\end{align}
Integration of \eqref{CellPbP5} on $(y_{\ell},y_{\ell+1})$ gives
\begin{equation}
\beta\left(P_2+\partial_y P_5\right)=\mathscr{C}_{\partial P_5,\ell}(t),\qquad \ell=1,\cdots, N.
\end{equation}
The continuity of flux in \eqref{CellPbP5} ensures that $\mathscr{C}_{\partial P_5,\ell}:=\mathscr{C}_{\partial P_5}$ does not depend on $\ell$, and hence
\begin{equation}
\partial_y P_5=\frac{1}{\beta}\mathscr{C}_{\partial P_5}-P_2.
\end{equation}
Averaging gives
\begin{equation}
-\sum_{\ell =1}^{N} \jump{P_5}_\ell=\mean{\frac{1}{\beta}}\mathscr{C}_{\partial P_5}-\mean{P_2},
\end{equation}
and hence $\mathscr{C}_{\partial P_5}=0$. It follows
\begin{equation}
P_2+\partial_y P_5=0.
\label{P2dP5}
\end{equation}

\section{Variational identities for the cell problems}\label{AppVarId}

A set of five useful variational identities deduced from the cell problems $(\mathscr{P}_1)-(\mathscr{P}_4)$. They are obtained by taking the average of ${\eqref{CellPbP3}\times P_1-\eqref{pb_cell_P1}\times P_3}$, ${\eqref{CellPbP4}\times P_1-\eqref{pb_cell_P1}\times P_4}$, ${\eqref{CellPbP3}\times \partial_t P_1-\partial_t\eqref{pb_cell_P1}\times P_3}$, ${\eqref{CellPbP3}\times \partial^2_{tt}P_1-\partial^2_{tt}\eqref{CellPbP2}\times P_3}$, and ${\eqref{CellPbP4}\times \partial_t P_1-\partial_t\eqref{pb_cell_P1}\times P_4}$, respectively and are given by:
\begin{align}
   \mean{\beta\,\partial_y P_3-\alpha\,P_1}-\sum_{\ell=1}^{N}\mathscr{m}_\ell(t)\,\dmean{P_1}_\ell=0,
\label{IRiii} \\[-4pt]
\mean{\beta\,\partial_y P_4}-\sum_{\ell=1}^{N}\mathscr{m}'_\ell(t)\,\dmean{P_1}_\ell=0,
\label{IRiv} \\[-4pt]
\mean{-\alpha\,\partial_t P_1}-\sum_{\ell=1}^{N}\mathscr{m}_\ell(t)\,\dmean{\partial_t P_1}_\ell=0,
\label{IRv} \\[-4pt]
\mean{-\alpha\,\partial^2_{tt} P_1}-\sum_{\ell=1}^{N}\mathscr{m}_\ell(t)\,\dmean{\partial^2_{tt} P_1}_\ell=0,
\label{IRvi} \\[-4pt]
-\sum_{\ell=1}^{N}\mathscr{m}'_\ell(t)\,\dmean{\partial_t P_1}_\ell=0.
\label{IRvii}
\end{align}

\vspace*{-20pt}
{
\bibliographystyle{RS}

\bibliography{sample}}

@article{ammariJoMP2023,
  title = {Transmission properties of time-dependent one-dimensional metamaterials},
  author = {Ammari, H. and Cao, J. and Hiltunen, E. O. and Rueff, L.},
  year = {2023},
  journal = {J. Math. Phys.},
  volume = {64},
  number = {12},
  pages = {121502},
  issn = {0022-2488},
  urldate = {2025-02-05}
}

@article{Galiffi2026,
	title = {Optical coherent perfect absorption and amplification in a time-varying medium},
	volume = {20},
	rights = {2026 The Author(s), under exclusive licence to Springer Nature Limited},
	issn = {1749-4893},
	pages = {163--169},
	number = {2},
	journal = {Nat. Photon.},
    year = {2026},
	publisher = {Nature Publishing Group},
	author = {Galiffi, Emanuele and Harwood, Anthony C. and Vezzoli, Stefano and Tirole, Romain and Alù, Andrea and Sapienza, Riccardo},
	date = {2026-02},
}

@article{Croenne2019,
  author = {Croënne, C. and Vasseur, J. O. and Bou Matar, O. and Hladky-Hennion, A.-C. and Dubus, B.},
  journal = {J. Appl. Phys.},
  title = {Non-reciprocal behavior of one-dimensional piezoelectric structures with space-time modulated electrical boundary conditions},
  year = {2019},
  issn = {1089-7550},
  number = {14},
  volume = {126},
  pages = {145108},
  publisher = {AIP Publishing},
}

@Article{Boutin1995,
  author    = {C. Boutin},
  journal   = {International Journal of Heat and Mass Transfer},
  title     = {Microstructural influence on heat conduction},
  year      = {1995},
  month     = {nov},
  number    = {17},
  pages     = {3181--3195},
  volume    = {38},
  publisher = {Elsevier {BV}}
}

@article{Huidobro2021,
  author = {P.A. Huidobro and M.G. Silveirinha and E. Galiffi and J.B. Pendry},
  journal = {Phys. Rev. Appl.},
  title = {Homogenization theory of space-time metamaterials},
  year = {2021},
  month = {jul},
  number = {1},
  pages = {014044},
  volume = {16},
  publisher = {American Physical Society (APS)},
}

@article{Huidobro2019,
  author = {Paloma A. Huidobro and Emanuele Galiffi and Sébastien Guenneau and Richard V. Craster and J. B. Pendry},
  journal = {Proc. Natl. Acad. Sci. U.S.A.},
  title = {Fresnel drag in space–time-modulated metamaterials},
  year = {2019},
  month = {nov},
  number = {50},
  pages = {24943--24948},
  volume = {116},
  publisher = {Proceedings of the National Academy of Sciences},
}

@article{Swinteck2015,
  author = {N. Swinteck and S. Matsuo and K. Runge and J. O. Vasseur and P. Lucas and P. A. Deymier},
  journal = {J. Appl. Phys.},
  title = {Bulk elastic waves with unidirectional backscattering-immune topological states in a time-dependent superlattice},
  year = {2015},
  month = {aug},
  number = {6},
  pages = {063103},
  volume = {118},
  publisher = {AIP Publishing},
}

@article{Cassedy1967,
  author = {E.S. Cassedy},
  journal = {Proc. IEEE},
  title = {Dispersion relations in time-space periodic media part II—Unstable interactions},
  year = {1967},
  number = {7},
  pages = {1154--1168},
  volume = {55},
  publisher = {Institute of Electrical and Electronics Engineers (IEEE)},
}

@article{Cassedy1963,
  author = {E.S. Cassedy and A.A. Oliner},
  journal = {Proc. IEEE},
  title = {Dispersion relations in time-space periodic media: Part I—Stable interactions},
  year = {1963},
  number = {10},
  pages = {1342--1359},
  volume = {51},
  publisher = {Institute of Electrical and Electronics Engineers (IEEE)},
}

@article{Salehi2022,
  author = {Salehi, Mohammadreza and Rahmatian, Pegah and Memarian, Mohammad and Mehrany, Khashayar},
  journal = {Opt. Express},
  title = {Frequency conversion in time-varying graphene microribbon arrays},
  year = {2022},
  issn = {1094-4087},
  number = {18},
  pages = {32061},
  volume = {30},
  publisher = {Optica Publishing Group},
}

@article{Nassar2020,
  author = {Hussein Nassar and Behrooz Yousefzadeh and Romain Fleury and Massimo Ruzzene and Andrea Alù and Chiara Daraio and Andrew N. Norris and Guoliang Huang and Michael R. Haberman},
  journal = {Nat. Rev. Mater.},
  title = {Nonreciprocity in acoustic and elastic materials},
  year = {2020},
  month = {jul},
  number = {9},
  pages = {667--685},
  volume = {5},
  publisher = {Springer Science and Business Media LLC},
}

@article{Lustig2018,
  author = {Lustig, Eran and Sharabi, Yonatan and Segev, Mordechai},
  journal = {Optica},
  title = {Topological aspects of photonic time crystals},
  year = {2018},
  issn = {2334-2536},
  number = {11},
  pages = {1390},
  volume = {5},
  publisher = {Optica Publishing Group},
}

@article{Touboul2024,
  author = {Touboul, Marie and Lombard, Bruno and Assier, Raphael C. and Guenneau, Sebastien and Craster, Richard V.},
  journal = {Proc. R. Soc. A},
  title = {High-order homogenization of the time-modulated wave equation: non-reciprocity for a single varying parameter},
  year = {2024},
  issn = {1471-2946},
  number = {2289},
  volume = {480},
  publisher = {The Royal Society},
}

@article{Nassar2017,
  author = {H. Nassar and X.C. Xu and A.N. Norris and G.L. Huang},
  journal = {J. Mech. Phys. Solids},
  title = {Modulated phononic crystals: Non-reciprocal wave propagation and Willis materials},
  year = {2017},
  pages = {10--29},
  volume = {101},
  publisher = {Elsevier BV},
}

@article{Ammari2022,
  author = {Ammari, Habib and Cao, Jinghao and Hiltunen, Erik Orvehed},
  journal = {Multiscale Model. Simul.},
  title = {NonReciprocal wave propagation in space-time modulated media},
  year = {2022},
  issn = {1540-3467},
  number = {4},
  pages = {1228--1250},
  volume = {20},
  publisher = {Society for Industrial & Applied Mathematics (SIAM)},
}

@article{bellisJotMaPoS2021,
  title = {Effective dynamics for low-amplitude transient elastic waves in a 1D periodic array of non-linear interfaces},
  author = {Bellis, Cédric and Lombard, Bruno and Touboul, Marie and Assier, Raphaël},
  year = {2021},
  journal = {J. Mech. Phys. Solids},
  volume = {149},
  pages = {104321},
  issn = {0022-5096},
  urldate = {2025-02-05}
}

@article{benjaziaWM2014,
  title = {Wave propagation in a fractional viscoelastic Andrade medium: diffusive approximation and numerical modeling},
  shorttitle = {Wave Propagation in a Fractional Viscoelastic Andrade Medium},
  author = {Ben Jazia, A. and Lombard, B. and Bellis, C.},
  year = {2014},
  journal = {Wave Motion},
  volume = {51},
  number = {6},
  pages = {994--1010},
  issn = {0165-2125},
  urldate = {2025-02-05}
}

@book{lurie2007introduction,
  title = {An introduction to the mathematical theory of dynamic materials},
  author = {Lurie, Konstantin A},
  volume = {15},
  year = {2007},
  publisher = {Springer}
}

@article{sanguinet2011homogenized,
  title = {The homogenized equations of motion for an activated elastic lamination in plane strain},
  author = {Sanguinet, WC},
  journal = {Z. Angew. Math. Mech.},
  volume = {91},
  number = {12},
  pages = {944--956},
  year = {2011},
  publisher = {Wiley Online Library}
}

@article{to2009homogenization,
  title = {Homogenization of dynamic laminates},
  author = {To, Hansun T},
  journal = {J. Math. Anal. Appl.},
  volume = {354},
  number = {2},
  pages = {518--538},
  year = {2009},
  publisher = {Elsevier}
}

@article{tien1958parametric,
  title = {Parametric amplification and frequency mixing in propagating circuits},
  author = {Tien, Ping K},
  journal = {J. Appl. Phys.},
  volume = {29},
  number = {9},
  pages = {1347--1357},
  year = {1958},
  publisher = {American Institute of Physics}
}

@article{cullen1958travelling,
  title = {A travelling-wave parametric amplifier},
  author = {Cullen, AL},
  journal = {Nature},
  volume = {181},
  number = {4605},
  pages = {332--332},
  year = {1958},
  publisher = {Nature Publishing Group UK London}
}

@book{craster2013,
  title = {Acoustic Metamaterials: Absorption, Cloaking, Imaging, Time-Modulated Media, and Topological Crystals},
  shorttitle = {Acoustic Metamaterials},
  author = {Craster, Richard V. and Guenneau, Sébastien},
  year = {2024},
  publisher = {Springer Series in Material Science (2nd Edition)},
  googlebooks = {uv4lQ0tQJtkC},
  isbn = {978-94-007-4813-2},
  langid = {english}
}

@article{galiffiA2022,
  title = {Photonics of time-varying media},
  author = {Galiffi, Emanuele and Tirole, Romain and Yin, Shixiong and Li, Huanan and Vezzoli, Stefano and Huidobro, Paloma A. and Silveirinha, Mário G. and Sapienza, Riccardo and Alù, Andrea and Pendry, J. B.},
  year = {2022},
  journal = {Adv. Photonics},
  volume = {4},
  number = {1},
  pages = {014002},
  publisher = {SPIE},
  issn = {2577-5421, 2577-5421},
  urldate = {2025-02-05}
}

@article{kimPRE2023,
  title = {Dynamics of time-modulated, nonlinear phononic lattices},
  author = {Kim, B. L. and Chong, C. and Hajarolasvadi, S. and Wang, Y. and Daraio, C.},
  year = {2023},
  journal = {Phys. Rev. E},
  volume = {107},
  number = {3},
  pages = {034211},
  publisher = {American Physical Society},
  urldate = {2025-02-05}
}

@article{kiorpelidisPRB2024,
  title = {Transient amplification in stable Floquet media},
  author = {Kiorpelidis, Ioannis and Diakonos, Fotios K. and Theocharis, Georgios and Pagneux, Vincent},
  year = {2024},
  journal = {Phys. Rev. B},
  volume = {110},
  number = {13},
  pages = {134315},
  publisher = {American Physical Society},
  urldate = {2025-02-05}
}

@article{craster2023mechanical,
  title={Mechanical metamaterials},
  author={Craster, Richard and Guenneau, S{\'e}bastien and Kadic, Muamer and Wegener, Martin},
  journal={Reports on Progress in Physics},
  volume={86},
  number={9},
  pages={094501},
  year={2023},
  publisher={IOP Publishing}
}

@article{mallejacPRA2023,
  title = {Scattering from time-modulated transmission-line loads: Theory and experiments in acoustics},
  shorttitle = {Scattering from Time-Modulated Transmission-Line Loads},
  author = {Malléjac, Matthieu and Fleury, Romain},
  year = {2023},
  journal = {Phys. Rev. Appl.},
  volume = {19},
  number = {6},
  pages = {064012},
  publisher = {American Physical Society},
  urldate = {2025-02-05}
}

@article{puJoSaV2024,
  title = {A Multiple scattering formulation for elastic wave propagation in space--time modulated metamaterials},
  author = {Pu, Xingbo and Marzani, Alessandro and Palermo, Antonio},
  year = {2024},
  journal = {J. Sound Vib.},
  volume = {573},
  pages = {118199},
  urldate = {2025-02-05}
}

@article{tessierbrothelandeAPL2023,
  title = {Experimental evidence of nonreciprocal propagation in space-time modulated piezoelectric phononic crystals},
  author = {Tessier Brothelande, S. and Croënne, C. and Allein, F. and Vasseur, J. O. and Amberg, M. and Giraud, F. and Dubus, B.},
  year = {2023},
  journal = {Appl. Phys. Lett.},
  volume = {123},
  number = {20},
  pages = {201701},
  issn = {0003-6951},
  urldate = {2025-02-05}
}

@article{caloz2019spacetime1,
  title = {Spacetime metamaterials—part I: general concepts},
  author = {Caloz, Christophe and Deck-Léger, Zoé-Lise},
  journal = {IEEE Trans. Antennas Propag.},
  volume = {68},
  number = {3},
  pages = {1569--1582},
  year = {2019},
  publisher = {IEEE}
}

@article{caloz2019spacetime2,
  title = {Spacetime metamaterials—Part II: Theory and applications},
  author = {Caloz, Christophe and Deck-Leger, Zoe-Lise},
  journal = {IEEE Trans. Antennas Propag.},
  volume = {68},
  number = {3},
  pages = {1583--1598},
  year = {2019},
  publisher = {IEEE}
}

@book{Bensoussan2011,
  title = {Asymptotic analysis for periodic structures},
  author = {Bensoussan, Alain and Lions, Jacques-Louis and Papanicolaou, George},
  volume = {374},
  year = {2011},
  publisher = {American Mathematical Soc.}
}

@book{Bakhvalov1989,
  author = {N. Bakhvalov and G. Panasenko},
  publisher = {Springer Netherlands},
  title = {Homogenisation: Averaging Processes in Periodic Media},
  year = {1989},
}

@article{Ammari2021,
  author = {Ammari, Habib and Hiltunen, Erik Orvehed},
  journal = {J. Comput. Phys.},
  title = {Time-dependent high-contrast subwavelength resonators},
  year = {2021},
  issn = {0021-9991},
  month = nov,
  pages = {110594},
  volume = {445},
  publisher = {Elsevier BV},
}

@article{Conca1997,
  author = {Carlos Conca and Muthusamy Vanninathan},
  journal = {SIAM J. Appl. Math.},
  title = {Homogenization of periodic structures via Bloch decomposition},
  year = {1997},
  month = {dec},
  number = {6},
  pages = {1639--1659},
  volume = {57},
  publisher = {Society for Industrial & Applied Mathematics (SIAM)}
}

@book{cioranescu1999,
  author = {Cioranescu, D. and Donato, P.},
  publisher = {Oxford University Press},
  title = {An Introduction to Homogenization},
  year = {1999},
  isbn = {9780198565536},
  lccn = {99033467},
  url = {https://books.google.co.uk/books?id=bggzAAAACAAJ},
}

@article{torrentPRB2018,
  title = {Loss compensation in time-dependent elastic metamaterials},
  author = {Torrent, Daniel and Parnell, William J. and Norris, Andrew N.},
  year = {2018},
  journal = {Phys. Rev. B},
  volume = {97},
  number = {1},
  pages = {014105},
  publisher = {American Physical Society},
  urldate = {2025-02-05}
}

@article{Touboul2024b,
  author = {Touboul, Marie and Lombard, Bruno and Assier, Raphael C. and Guenneau, Sebastien and Craster, Richard V.},
  journal = {Proc. R. Soc. A},
  title = {Propagation and non-reciprocity in time-modulated diffusion through the lens of high-order homogenization},
  year = {2024},
  issn = {1471-2946},
  month = nov,
  number = {2301},
  volume = {480},
  publisher = {The Royal Society},
}

@article{wenCP2022,
  title = {Unidirectional amplification with acoustic non-Hermitian space-time varying metamaterial},
  author = {Wen, Xinhua and Zhu, Xinghong and Fan, Alvin and Tam, Wing Yim and Zhu, Jie and Wu, Hong Wei and Lemoult, Fabrice and Fink, Mathias and Li, Jensen},
  year = {2022},
  journal = {Commun. Phys.},
  volume = {5},
  number = {1},
  pages = {1--7},
  publisher = {Nature Publishing Group},
  issn = {2399-3650},
  urldate = {2025-02-05},
  copyright = {2022 The Author(s)},
  langid = {english}
}

@article{zhuAPL2020,
  title = {Non-reciprocal acoustic transmission via space-time modulated membranes},
  author = {Zhu, Xiaohui and Li, Junfei and Shen, Chen and Peng, Xiuyuan and Song, Ailing and Li, Longqiu and Cummer, Steven A.},
  year = {2020},
  journal = {Appl. Phys. Lett.},
  volume = {116},
  number = {3},
  pages = {034101},
  issn = {0003-6951},
  urldate = {2025-02-05}
}

@article{zhuPRB2020,
  title = {Tunable unidirectional compact acoustic amplifier via space-time modulated membranes},
  author = {Zhu, Xiaohui and Li, Junfei and Shen, Chen and Zhang, Guangyu and Cummer, Steven A. and Li, Longqiu},
  year = {2020},
  journal = {Phys. Rev. B},
  volume = {102},
  number = {2},
  pages = {024309},
  publisher = {American Physical Society},
  urldate = {2025-02-05}
}

@article{ammari_spacetime_2025,
  title = {Space–time wave localization in systems of subwavelength resonators},
  journal = {Proc. R. Soc. A},
  author = {Ammari, Habib and Hiltunen, Erik Orvehed and Rueff, Liora},
  year = {2025},
  volume = {480},
  number = {2289},
  pages = {20240177},
}

@article{ammari_scattering_2024,
  title = {Scattering from time-modulated subwavelength resonators},
  volume = {480},
  pages = {20240177},
  number = {2289},
  journal = {Proc. R. Soc. A},
  author = {Ammari, Habib and Cao, Jinghao and Hiltunen, Erik Orvehed and Rueff, Liora},
  year = {2024},
}

@article{Simon1960,
  author = {J.-C. Simon},
  journal = {IEEE Trans. Microw. Theory Tech.},
  title = {Action of a progressive disturbance on a guided electromagnetic wave},
  year = {1960},
  month = {jan},
  number = {1},
  pages = {18--29},
  volume = {8},
  publisher = {Institute of Electrical and Electronics Engineers (IEEE)},
}

@article{Oliner1961,
  author = {A.A. Oliner and A. Hessel},
  journal = {IEEE Trans. Microw. Theory Tech.},
  title = {Wave propagation in a medium with a progressive sinusoidal disturbance},
  year = {1961},
  month = {jul},
  number = {4},
  pages = {337--343},
  volume = {9},
  publisher = {Institute of Electrical and Electronics Engineers (IEEE)},
}

@article{Koukouraki2025,
  title = {Floquet scattering of shallow water waves by a vertically oscillating plate},
  journal = {Wave Motion},
  volume = {136},
  pages = {103530},
  year = {2025},
  issn = {0165-2125},
  author = {Magdalini Koukouraki and Philippe Petitjeans and Agnès Maurel and Vincent Pagneux}
}

@article{Darche2025,
  author = {Darche, Michaël and Assier, Raphaël and Guenneau, Sébastien and Lombard, Bruno and Touboul, Marie},
  journal = {C. R. Méc.},
  title = {Scattering of transient waves by an interface with time-modulated jump conditions},
  year = {2025},
  issn = {1873-7234},
  month = jul,
  number = {G1},
  pages = {923--951},
  volume = {353},
  publisher = {Cellule MathDoc/Centre Mersenne},
}

@book{Strickwerda2004,
  author = {Strickwerda, J. C.},
  publisher = {SIAM},
  title = {Finite-Difference Schemes and Partial Differential Equations},
  year = {2004},
}

\end{document}